\documentclass[NPG]{copernicus}

\newcommand{\etal}{{et~al. }}

\newcommand{\G}{\mathcal{G}}

\newcommand{\smax}{s_{\mathrm{max}}}
\newcommand{\smin}{s_{\mathrm{min}}}

\newcommand{\ave}[1]{\left\langle#1 \right\rangle}

\newcommand{\flabel}[1]{\label{fig:#1}}
\newcommand{\fref}[1]{Fig.~\ref{fig:#1}}
\newcommand{\Fref}[1]{Figure ~\ref{fig:#1}}
\usepackage{ulem}
\usepackage{subfig}
 \usepackage{color}

\usepackage{comment}
\usepackage{graphicx}
\newcommand{\AD}[1]{{\bf ***AD #1 ***}}         %Alvaro Corral
\begin{document}

%stechmann2011stochastic
%yano2012self
%veneziano2012scaling
%olami1992temporal
%schertzer1994multifractal
%hooge1994mulifractal
%deidda2000rainfall,
  %title={Rainfall downscaling in a space-time multifractal framework},
 %lovejoy1995multifractals,
  %title={Multifractals and rain},

\title{
%Scale Invariant Properties for Medium Resolution Local Rain Data \\
%or 
Scale Invariant Events and Dry Spells for Medium Resolution Local Rain Data% \\
%\sout{Power Laws and Scaling of Rain Events and Dry Spells in the Catalonia Region}
%Scaling of Dry Spells and Event Sizes in Non-high Resolution Rain Data
%Scaling properties of Dry Spells and Event sizes for Low Resolution Local Rain DataScaling Properties %of Dry Spells and Event Sizes for 
%on Medium Resolution Local Rain Data
}

%\author{Anna Deluca and \'Alvaro Corral}

\author[1,2]{Anna Deluca}
\author[1]{ \'Alvaro Corral}
%\author[]{NAME}

\affil[1]{Centre de Recerca Matem\`atica, Edifici Cc, Campus Bellaterra, 
E-08193 Bellaterra (Barcelona), Spain}
\affil[2]{Departament de Matem\`atiques, Universitat Aut\`onoma de Barcelona, 08193 Bellaterra (Barcelona), Spain}

%% The [] brackets identify the author to the corresponding affiliation, 1, 2, 3, etc. should be inserted.

\runningtitle{
Scale Invariant Events and Dry Spells for Medium Resolution Local Rain Data
%Scale Invariant Properties in Medium Resolution Local Rain Data
%Scaling properties
% of Dry Spells and Event sizes for 
%on Medium Resolution Local Rain Data
}

\runningauthor{A. Deluca and \ A. Corral}

\correspondence{A. Deluca (adeluca@crm.cat), A. Corral (acorral@crm.cat)}

\received{}
\pubdiscuss{} %% only important for two-stage journals
\revised{}
\accepted{}
\published{}

%% These dates will be inserted by the Publication Production Office during the typesetting process.

\firstpage{1}

\maketitle

\begin{abstract}
%\sout{
%We analyze the statistics of rain-event sizes, rain-event durations,
%and dry-spell durations in a network of 20 rain gauges scattered
%in an area situated %%in NE Spain.
%close to the NW Mediterranean coast.
%We carefully neglect the parts of the data affected by errors.
%Power-law distributions emerge clearly for the dry-spell durations,
%with an exponent around $1.50 \pm 0.05$,
%although for event sizes and durations the power-law ranges
%are rather limited, in some cases.
%Deviations from power-law behavior are attributed
%to finite-size effects.
%A scaling analysis helps to elucidate the situation,
%providing support for the existence of scale invariance
%in these distributions.
%It is remarkable that rain data of not very high resolution
%yield findings in agreement with %% the recently proposed
%self-organized critical phenomena.
%Our research also shows how it is not necessary
%to rely on rain data of very high resolution
%to obtain agreement with the self-organized critical.}
%%
%\sout{This contribution} 
We analyze distributions of rain-event sizes, rain-event durations,
and dry-spell durations 
%\sout{distributions} 
for data obtained from a network of 20 rain gauges scattered
%in an area situated
%in NE Spain.
%close to
%\sout{in the NW Mediterranean coast.}
in a region of the NW Mediterranean coast.
%We carefully neglect the parts of the data affected by errors.
While power-law distributions model the dry-spell durations
%Power-law distributions emerge clearly for the dry-spell durations,
with a common exponent %around
$1.50 \pm 0.05$, density analysis is inconclusive for
event sizes and  event durations, %\sout{for some datasets.}
 due to finite size effects.
%results not show clear evidence by
%for all the studied cases.
%although for event sizes and durations the power-law ranges
%are in some cases limited.%rather limited, in some cases.
%Deviations from power-law behavior are attributed to finite-size effects.
However, we present  alternative evidence of
%In order An scaling analysis supports
%We A scaling analysis %helps to elucidate the situation,
%providing support for
the existence of scale invariance
in these distributions 
 by means of different data collapses of the distributions.
%\sout{supporting the proposed statistical model.}
%It is remarkable that rain data of not very high resolution
%yield findings in agreement with %% the recently proposed
%self-organized critical phenomena.
These results 
%\sout{show that scaling properties can also be observed for medium resolution data,}
 are in agreement with the expectations from the
Self-Organized Criticality paradigm,
and demonstrate that scaling properties of rain events and dry spells 
can also be observed for medium resolution  rain data.
%that the use of very high resolution data for the observation
%of scaling properties can be relaxed. % the unnecessity , already show as
%Our research also shows how it is not necessary
%it is not necessary
%to rely on rain data of very high resolution
%to obtain agreement with the self-organized critical
%view of rainfall.
%\hilight{REVISION NEEDED } \hilightgreen{1}
%{\bf NO DECIMOS NADA DE SOC!!! 
%PERO EN LAS CONCLUS SOC ES IMPORTANTE!!!}
\end{abstract}

%\begin{comment}

\introduction
%% \introduction[modified heading if necessary]

%\AD{General: change notation? remove $s$ subscript on $P(s)$?}

%\AD{INTRO: What to change to satisfy Referee 2:

%- Explain better relationship between power laws and SOC

%- Explain similarities and differences between SOC and Multrifractals,
%as an alternative mechanism that also predict power-law distributions

%- Discuss the effects of the quality of the data on the fact that high resolution
% measuring devices are necessary to obtain sufficient large range of power-law behavior. 
%ESTO MEJOR A LAS CONCLUSIONES!!!
%}

%%-----------------------------------
%{%\bf 1. INTRODUCIENDO SUPRISING REGULARITIES}

The complex atmospheric processes related to precipitation and convection 
arise from the cooperation
of diverse non-linear mechanisms with different temporal and 
spatial characteristic scales. 
Precipitation 
%\sout{Rain occurrence involves}
 combines, for instance, the $O(100\mu m)$ microphysics effects as evaporation with 
$O(1000km)$ planetary circulation of masses and moisture.
% phenomena from the microphysics scale ...
%\sout{\bf  An every-day event like precipitation combines, 
%for instance, the $O(100\mu m)$ microphysics effects as evaporation with 
%$O(1000km)$ planetary circulation of masses and moisture}.
% \AD{Lo he cambiado bastante, lo podriamos conservar} \sout{Moreover,} 
 Rain fields also presents  high  spatial and temporal intermittency 
%\sout{,  i.e., most of {\bf the time} it does not rain, and when it does,} 
 as well as extreme variability, in such a way that their 
intensity cannot be characterized by its mean value 
%\sout{due to its high variability} 
\citep{bodenschatz2010can}. 
Despite the complexity %\sout{and richness}
 of 
%\sout{individual events} 
the processes involved, surprising statistical regularities have been found:   numerous geometric and radiative properties of clouds present clear scaling or multiscaling behavior \citep{lovejoy1982area, cahalan1989fractal, peters2009mesoscale, wood2011distribution}; 
%\sout{or}
 also, raindrop arrival times and raindrop sizes, 
%\sout{which also} 
are well characterized by power-law distributions over several of orders of magnitude \citep{olsson1993fractal,lavergnat2006stochastic}. 
%\hilight{chequear ultima frase}
%{\bf  ESTA FRASE NO ACABA DE ESTAR BIEN REDACTADA ???} 
%Despite the complexity  of individual events, surprising statistical regularities have been found for different observables: {\bf numerous geometric and radiative properties of clouds, raindrop arrival times and raindrop sizes, for just cite some examples, present clear scaling or multi-scaling statistics \citep{lovejoy1982area, cahalan1989fractal, peters2009mesoscale, wood2011distribution},  and are all well characterized by long-tailed distributions over several orders of magnitude \citep{olsson1993fractal,lavergnat2006stochastic}}.

%{\bf  Cloud size and other geometric properties present clear scaling relationships and power-law distributions  \citep{lovejoy1982area, cahalan1989fractal, peters2009mesoscale, wood2011distribution}, as well as,  raindrop arrival  or size distribution, which also are well characterized by long-tailed distributions over several of orders of magnitude \citep{olsson1993fractal,lavergnat2006stochastic}}.

{%\bf Numerous geometric and radiative properties of clouds present clear scaling or {\bf multiscaling} statistics  Cloud size and other geometric properties present clear scaling relationships and power-law distributions  \citep{lovejoy1982area, cahalan1989fractal, peters2009mesoscale, wood2011distribution}, as well as,  raindrop arrival  or size distribution, which also are well characterized by long-tailed distributions over several of orders of magnitude \citep{olsson1993fractal,lavergnat2006stochastic}}.

%lovejoy1982area
%cahalan1989fractal
%olsson1993fractal
%lavergnat2006stochastic
%peters2009mesoscale
%wood2011distribution

%\AC{Ref needed!!!} \AD{LIST OF STATISTICAL ReGuALRITIES, INCLUDING POWER-LAWS}

%%-----------------------------------
%{\bf 2. SOC intro}

The concept of  self-organized criticality (SOC) aims for explaining the origin of
the emergence of structures across many different 
spatial and temporal scales 
%% power-law distributed event sizes
in a broad variety of systems 
\citep{Bak_book,Jensen,Sornette_critical_book,Christensen_Moloney}.
Indeed, it has been found that for diverse phenomena
that take place intermittently, in terms of 
%bursty episodes of activity 
bursts of activity
interrupting larger quiet periods, 
the size $s$ of these bursty events or avalanches follows a power-law
distribution,
\begin{equation}
%\begin{center}
\centering
P(s) \propto \frac 1 {s^{\tau_s}},
%\end{center}
\label{powerlaw}
\end{equation}
over a certain range of $s$,
with $P(s)$ the probability density of the event size
and $\tau_s$ its exponent
(and the sign $\propto$ indicating proportionality).
The size $s$ 
%\sout{is}
 can be understood as a
% \sout{more or less direct}
 measure
of energy dissipation.
If durations of events are measured,  a
power-law distribution also holds.
These power-law distributions provide an unambiguous proof
of the absence of characteristic scales within the avalanches, as power laws are the only
fully scale-invariant functions \citep{Christensen_Moloney}.

%Power-law distributions
%signal the absence of characteristic scales %% event sizes 
%%in the occurrence of events
%%in the system
%\citep{Christensen_Moloney}.
%{\AD demasiadas autocitas cita Corral_Lacidogna quitada}
%,Corral_Lacidogna}.
%\AD{cita Corral quitada}

The main idea behind SOC is the recognition that such scale invariance
is achieved because of the existence of a non-equilibrium continuous phase transition
whose critical point is an attractor of the dynamics %. %% of the system.
% Anyadir aqui citas!!!!!!
\citep{Tang_Bak_prl,Dickman_pre98,Dickman_bjp}.
When the system settles at the critical point, 
scale invariance and power-law behavior are ensured,
as these peculiarities are the defining characteristics of critical phenomena
\citep{Christensen_Moloney}. 
%\sout{This will be our definition},  
%\sout{This is the definition that we adopt,} 
%\AD{\hilight{Te gusta Jefe? vale...}} \hilightgreen{2}
 Although sometimes SOC is understood in a broader sense, as the spontaneous
emergence of scale invariance, we will follow the previous less-ambiguous definition. 
%\sout{
The concept of SOC has had big impact in the geosciences, in particular  earthquakes \citep{Bak_book,Sornette_sornette}, 
landslides and rock avalanches \citep{Malamud_hazards}, 
%\sout{
%volcanic eruptions \citep{Malamud_hazards,Lahaie},} 
or  
forest fires \citep{Malamud_science}. %\sout{, among others, which} 
%\sout{ and even the extinction of biological species \citep{Sneppen_PNAS}},
Due to the existence of power-law distributed events in them, 
these systems have been proposed as 
%%can be considered as 
realizations of SOC in the natural world.%}

The SOC perspective
has also been applied to rainfall,
% \sout{, which} 
looking at precipitation as an avalanche process,  
and paying attention to the properties of these avalanches, called rain events. %\sout{different observables \sout{that} {\bf than} the rain rate.}
 The first 
%%results 
%%viewing rainfall this way are 
works following this approach are
those of \cite{Andrade} and 
%\cite{Peters_prl} \citep{Peters_pre,Peters_jh}.
Peters \etal (2002; Peters and Christensen 2002, 2006). 
%who studied rainfall as an avalanche process.
 These authors  defined, independently, a rain event as 
the sequence of rain occurrence with rain rate (i.e., the activity) always greater than zero.
Then, the focus of the SOC approach is not on the total amount of rain recorded in a 
fixed time period (for instance,  one hour, one day, or one month), but on the rain event, which is what defines in each case the time period of rain-amount integration. In this way, the event size is the total amount of rain collected during the duration of the event.

%{\bf 3. MORE DETAILS ON EARLY RESULTS, ANDRADE Y OLE, Universality paper}

Andrade \etal  studied long-term daily local (i.e., zero-dimensional)
rain records
from weather stations in Brazil, India, Europe, and Australia,
with observation times ranging from a 
decade to a century approximately,
with detection threshold 0.1 mm/day.
Although the dry spells (the times between rain events) seemed to follow in some case 
a steep power-law distribution, the rain-event size distributions were not reported, 
and therefore the connection between SOC and rainfall could not really  be 
checked.
 Later, Peters \etal analyzed 
%\sout{very} 
high resolution  rain
data from a vertically pointing Doppler radar situated
in the Baltic coast, which provided 
%\sout{rain} 
rates at an altitude between
250 m and 300 m above sea level, covering an area of 70 m$^2$, with detection threshold 0.0005 mm/hour 
%\AD{This is not rain! Creo que asi ya esta salvado!} 
and temporal resolution of 1 minute.  Power-law  distributions for event sizes and for dry-spells durations 
%%was obtained for
over several orders of magnitude were reported, with exponents $\tau_s \simeq \tau_q \simeq 1.4$. 
For the event-duration distribution the results were unclear, 
although a power law with an exponent $\tau_d\simeq 1.6$  was fit to the data.

 More recently, 
a study covering 10 sites across different climates  has checked the universality of rain-event statistics 
using rain data from optical gauges
\citep{Peters_Deluca}.
%%which also gives 
%\sout{ extending the support to the SOC hypothesis in rainfall.}
The %%rain 
data had a resolution of $0.2$ mm/hr, collected at 
intervals of 1 minute. 
The results showed unambiguous power-law distributions of
event sizes, with  apparent  universal exponents  
$\tau_s=1.17 \pm 0.03$,
extending the support to the SOC hypothesis in rainfall.
 Power laws distributions were also found for the dry spell durations, but
for event durations %\sout{and dry spell durations}
the behavior was not so clear.
%\sout{Frase reordenada!!}

%///////////////////////////////////

%{\bf 3. Diferentes propuestas para approach differentes}

%%-----------------------------------
%{\bf 4. Multifractal intro and comparison}
%* Clouds as a fractal

%\sout{However}
Nevertheless, scale-free distributions of the observables are insufficient 
evidence for SOC dynamics, as there are many alternative mechanisms 
%%for its production 
 of power-law genesis
\citep{Sornette_critical_book,Dickman_rain}.
 In other words, SOC implies power laws, 
but the reciprocal is not true, power laws are not a guarantee of SOC.
%%For instance, 
In particular, the multifractal approach can  also reproduce scale invariance, 
%%them, 
but  using different observables. 
 When applied to rainfall,
this approach 
 focus on the rain rate field, which is hypothesized to have multifractal support as a result from a multiplicative cascade process.  
% \sout{This} 
 From this point of view, %{\bf ,} %\sout{has lead to}  
alternative
statistical models %{\bf as well as} 
new forecasting and downscaling methods
 have emerged
\citep{lovejoy1995multifractals,deidda2000rainfall,veneziano2012scaling}. %\AD{\hilight{cambiar}}
  %title={Multifractals and rain}, [cite general lovejoy]
%and to the development of a theoretical framework explain these properties 

 In general, one can distinguish between
continuous and within-storm multifractal analysis. 
The first one considers the whole rain rate time series 
(including dry spells), while the second  one analyzes just rain rate time series within storms. This requires a storm definition, which usually contains dry periods too, but with duration  smaller than a certain threshold. 
%%\citep{veneziano2012scaling}.  %usually allows dry periods with duration smallest than a certain period.
The connections between SOC and the multifractal approach are still an open question,
despite some seminal works
\citep{olami1992temporal, schertzer1994multifractal,hooge1994mulifractal}.
 We expect that these connections could be developed more in depth
from the within-storm multifractal approach, which presents more similarities with the SOC one;
however, such an ambitious goal is beyond the scope of this article.
 %\AD{\hilight{Move to conclusions?? no cal}}
%\AD{DECIR MAS SOBRE ESTO EN LAS CONCLUSIONES, COMO QUE GUAI QUE SERIA
%QUE ESTUVIERAN MEJOR CONECTADAS Y CUAN GRANDE SERIA EL BENEFICIO MUTUO DE ESA CONEXION}

%{\bf 5. Y porque SOC? incluso hay alternativas a los eventos.. Nature Phys + Universalidad }

%%As there are alternative mechanisms which produce power-laws, a \sout{A} 
Coming back to the problem of SOC in rainfall,
a  more direct approach was undertaken by
 %\sout{followed by}
 \cite{Peters_np}.
% \sout{who} 
 They analyzed
satellite estimates of rain rate and
vertically integrated (i.e., column) water vapour content in grid points
covering the tropical oceans
(with a $0.25^\circ$ spatial resolution in latitude and longitude) from the Tropical Rainfall Measuring Mission,
%(from 2000 to 2005).
and they
 found a sharp increase of the rain rate
when a threshold value of the water vapor was reached,
%%analogous to {\bf the obtained in } 
 in the same way as in
critical phase transitions.
Moreover, these authors showed that most of the time
the state of the system was close to the transition point
(i.e., most of the measurements of the water vapor correspond
to values near the critical one), 
providing %%{\bf not only} 
%\sout{direct}
% {\bf  real, concrete ???????}  
%\sout{\bf straightforward}
%\sout{\bf conclusive}
 convincing %\hilight{la que decidas en la proxima se queda.
%Venga, me ganas por aburrimiento...} \hilightgreen{3}
observational support of the validity of 
SOC theory in  rainfall. 
%%atmospheric convection {\bf and a better characterization of the \sout{rainfall} phenomena.}
Further, they connected these ideas with the classical concept of 
%%atmospheric
quasi-equilibrium for atmospheric convection 
\citep{Arakawa_schubert}, allowing the application of the SOC ideas in cloud resolving model development \citep{stechmann2011stochastic}. %\AD{Cita Yun-Ichi sacada}%,yano2012self}.
%yano2012self[Stechmann and Neelin][Jun-Ichi].
Remarkably, as far as we know, an analogous result
has not been found in other claimed
SOC natural systems, as earthquakes or
forest fires;
 this would imply that the result of Peters and Neelin
is the first unambiguous proof of SOC in these systems.

%{\bf 6. What is missing to be known? which problems??}

 In any case, the existence of SOC in rainfall
%{\bf
 posses many {questions}.
%}. 
%\sout{is still a not fully solved issue} \sout{remains an open question}. 
 As we have seen, the number of studies addressing this 
%\hilightred{\sout{question}} {\bf matter}  \hilightgreen{4} 
is  rather limited, 
mostly due to 
the supposed requirement that the data has to be of very high time and rate resolution.
%\sout{limitations on the event definition, which highly depends on the data resolution (??????).}
Moreover, 
 testing further the critical dependence of rain rate on column water vapor (CWV), as seen 
in \cite{Peters_np}, is nonviable for local data due to current problems of the microwave radiometers at hight CWV values %with the current microwave radiometers
% due to problem with 
%probably because of the wet window problem and the problems the MWR can have at high CLW values. This makes it impossible to test the power-law relationship at high CWV, as seen in Peters and Neelin
%OJO!! FRASE NO COMPLETA!!!!!
%\sout{ testing the critical dependence of rain rate on water vapor %\hilight{column?} \hilightred{is not possible }with 
%\sout{in-situ (??????)}
 %local data, as %\AD{\hilight{the disdrometers cannot measure}} 
 %water vapor content
%\sout{is not available} 
%when it rains
%, making very difficult to test further its relationship with precipitation} 
\citep{ holloway2010temporal}. % \hilightgreen{5 ojo!!!}
 Finally, the kind of data analyzed by Peters and Neelin
is completely different to the data employed in the 
studies yielding power-law distributed events
\citep{Peters_prl,Peters_Deluca}, 
 so, direct comparison between both kinds of approaches is not possible.
%\sout{\bf making difficult comparison.}  \AD{\hilight{pero nosotros lo hacemos }} \AD{\hilight{y ellos tb miden
%power laws con exponentes 1.2?!
%Cuales son esos resultados? Con datos de la TRM miden distribuciones
%de eventos???}}
%\sout{so, both approaches remain somewhat disconnected.}

%Power-law size distributions alloconnection to be established:
%if SOC systems show distributions that are power law, 
%and the system under study displays a power-law distribution,
%there exists the possibility that the system is a SOC system, 
%although alternative mechanisms could explain the power law
%\citep{Sornette_critical_book,Dickman_rain} \AD{ADD about multiscaling mechanisms}.

%{\bf 7. What we do and why in this paper.}

The goal of this paper is 
to extend the evidence for SOC in rainfall,
%%not to solve this complex puzzle but
%%to contribute to %%to fill this gap,
%to extend the evidence for SOC in rainfall
%studying a climatology that can be considered
%as a link between the Baltic-sea case analyzed by Peters \etal
%and the tropical oceans of Peters and Neelin.
%looking for more evidence of scale-invariant rain-event distributions
%in sites with climatic characteristics distinct from
%the case considered by Peters \etal
%\cite{Peters_prl,Peters_pre,Peters_jh},
%and analyze %%the possible existence or not of a ``universal''.
%the nature of the differences and the possible similarities between them.
%to study the applicability of the SOC view of rainfall to the Mediterranean climatology.
studying the applicability of this paradigm 
%{\bf
 when the rain data available is not of high resolution.
%\AD{es este realmente nuestro goal? Nostamal, no???}
%%to the Mediterranean climatology.
With this purpose, we perform an in-depth analysis
of local rainfall records in 
a representative region of the Northwestern Mediterranean.
%\sout{As a by-product, we will show how non-high resolution rain data
%can \sout{still} also contain valuable information for \sout{SOC} scaling and SOC analysis.}
For this lower (in comparison with previous studies)
%\sout{In this smaller} 
resolution,
the range in which the power-law  holds can be
 substantially decreased.
%%{\bf be shorter} 
%\sout{be very short}. 
This may require the application of  
%\sout{the very} 
more refined 
fitting techniques and 
%\sout{to the}
 scaling methods. 
Thus, as a by-product,
we explore different scaling forms and develop a collapse method 
based on minimizing the distance between distributions  that also gives an estimation of 
the power-law exponent.
 With these tools will be able to establish the existence of
scale-invariant behavior in the medium resolution rain data analyzed.

%{\bf
 We proceed as follows: Section 2 describes the data used in the present analysis
%%and introduces {\bf the rain events and the dry spells}.
%{\bf
 and defines
%}
 the rain event, 
%{\bf
 its size and duration, and the dry spell.
%}.
Section 3
%{\bf 
shows
%}
 the corresponding probability densities and
%\sout{We {\bf also}} 
describes and applies an accurate fitting technique for evaluating 
the power-law existence.
Section 4 introduces 
two collapse methods (parametric and non-parametric) 
%{\bf
 in order to establish the fulfillment of scaling,
independently of power-law fitting.
%}
%%are presented in Sec 3.4 and Sec. 3.5. 
Discussion and conclusions are presented in section 5. 

%\AD{\hilight{rehacer al final.. ya estaria, no??}}

%BUFA, ESTO MEJOR ESCRIBIRLO CUANDO ESTE CLARO LO DE LAS SECCIONES,
%Y NO HACE FALTA ENTRAR EN SUBSECCIONES.
}

\section{Data and Definitions}

\subsection{Data} %\sout{Rain database} \hilight{Measurements}  \hilightgreen{6??? Por que measurements??? Nosotros no medimos nada...}}

%\AD{Data: What to change to satisfy Referee 2:

%- Not clear in order to make the datasets more manageable...zeros...

%- Discuss the effects of the quality of the data on the fact that high resolution
% measuring devices are necessary to obtain sufficient large range of power-law behavior. 

%-  with longitudes and latitudes ranging from 
%1$^\circ$ 10' 51'' to 3$^\circ$ 7' 35'' E  and from
%41$^\circ$ 12' 53'' to 43$^\circ$ 25' 40'' N respectively
%}

We have analyzed 20 stations in Catalonia (NE Spain)
from the database maintained by the
Ag\`encia Catalana de l'Aigua (ACA, {\tt http://aca-web.gencat.cat/aca}).
These data come from a network of rain gauges, 
called SICAT (Sistema Integral del Cicle de l'Aigua al Territori, 
formerly SAIH, Sistema Autom\`atic d'Informaci\'o Hidrol\`ogica),
used to monitor the state of the drainage basins
%of this region.
%The inland basins are those comprising the rivers
%which 
of the rivers that are born and die in the Catalan territory. 
%%%\citep{Llasat_aca}.
The corresponding sites are listed in Table \ref{tableone}
and have longitudes and latitudes ranging from 
1$^\circ$ 10' 51'' to 3$^\circ$ 7' 35'' E  and from
41$^\circ$ 12' 53'' to 43$^\circ$ 25' 40'' N.
%,together with their latitude and longitude.%;
%a map is also provided in  \fref{map}.
All datasets cover a time period
starting on January 1st, 2000, at 0:00,
and ending either on June 30th or on July 1st, 2009 
(spanning roughly 9.5 years),
except the Cap de Creus one,
which ends on June 19th, 2009.
%\sout{The same database, although for a different time period, was
%also used in \cite{Llasat_aca}.}

\begin{table*}[ht]
\caption{
Characteristics of all the sites for the 9-year period 2000-2008. 
%and
%%Longitud E 1deg 10 min 51 sec to 3 deg 7 min 35 sec
%%Latitud N 41 deg 12 min 53 sec // 43 deg 25 min 40 sec
%1$^\circ$ to 3$^\circ$ $E$ and 41$^\circ$ to 42$^\circ$ $N$ respectively.
%%Longitud E 1deg 10 min 51 sec // 3 deg 7 min 35 sec
%%Latitud N 41 deg 12 min 53 sec // 43 deg 25 min 40 sec
Every site is named by the corresponding river basin or subbasin
 (the municipality is included in ambiguous cases);
Ll. stands for Llobregat river.
$f_M$ is the fraction  (in \%) of missing records (time missing divided by total time);
$f_D$ is the fraction  (in \%)  of discarded times;
$f_r$ is the fraction   (in \%)  of rainy time 
(time with $r>0$ divided by total undiscarded time,
for a time resolution $\Delta t=5$ min);
a. rate is the annual rain rate  in mm/yr, calculated only over undiscarded times;
c. rate is the rain rate  in mm/hr conditioned to rain, i.e., 
calculated over the (undiscarded) rainy time;
$N_s$ is the number of rain events and $N_q$ the number of dry spells
%{\bf 
(the differences between $N_s$ and $N_q$ are due to the missing records);
%}
the rest of symbols are explained in the text.
 $\langle s\rangle$ is measured in mm, and 
$\langle d\rangle$ and $\langle q\rangle$ in min.
%%The differences between $N_s$ and $N_q$ are due to the missing records.
Sites are ordered by increasing annual rate.
The table shows a positive correlation between $f_r$, the annual rate, 
$N_s$ and $N_q$, and that these variables are negatively correlated
with $\ave{q}$. % and $\ave{q^2}/\ave{q}$.
In contrast, the rate conditioned to rain is roughly constant, 
taking values between 3.3 and 3.8 mm/hr.}
  \label{tableone}
\vskip4mm
%\hspace{-1.65cm}
\center
\begin{tabular}{l r r r r c r r r r r }
\tophline
\large{Site}  & $f_M$ & $f_D$ & $f_r$ & a. rate & c. rate & $N_s$ & $N_q$ & $\ave{s}$ & $\ave{d}$ & $\ave{q}$  \\
   %& \%  & \%  & \%  & mm/yr   & mm/hr    &       &       & mm      & min       & min       \\
\middlehline
1        Gai\`a   &   0.08  &  3.71  & 1.6  & 470.9 & 3.3 &  5021 &  5014 & 0.81 &      14.9 &     894.  \\
2                     Foix   & 0.07  &  3.38  & 1.6  & 500.6 & 3.6 &  4850 &  4844 & 0.90 &      15.0 &     929.  \\
3 Baix Ll. S.J. Desp\'{\i}   &  0.07  &  2.28  & 1.7  & 505.8 & 3.3 &  5374 &  5369 & 0.83 &           15.0 &   847.  \\
4                   Garraf   &  0.09  &  3.30  & 1.6  & 507.8 & 3.7 &  4722 &  4716 & 0.94 &     15.2 &     956.  \\
5     Baix Ll. Castellbell   &  0.06  &  2.81  & 1.7  & 510.7 & 3.4 &  4950 &  4947 & 0.90 &      15.8 &      914. \\
6           Francol\'{\i}    &  0.44  & 13.37  & 1.8  & 528.2 & 3.4 &  4539 &  4540 & 0.91 &      16.1 &     887.  \\
7          Bes\`os Barcelona &  0.15  &  4.17  & 1.7  & 531.8 & 3.5 &  4808 &  4803 & 0.95 &      16.2 & 928.  \\
8         Riera de La Bisbal &  0.07  &  3.66  & 1.6  & 540.0 & 3.8 &  4730 &  4724 & 0.99 &     15.8 &      950. \\
9          Bes\`os Castellar &  4.34  & 13.59  & 2.0  & 633.3 & 3.6 &  4918 &  4970 & 1.00 &        16.9 &     806.  \\
10            Ll. Cardener   &  0.07  &  3.33  & 2.1  & 652.4 & 3.5 &  6204 &  6197 & 0.92 &        15.7 &     723.  \\
11                 Ridaura   &0.12  &  2.41  & 2.0  & 674.2 & 3.8 &  5780 &  5774 & 1.02  &    16.1 &     784. \\
12                  Dar\'o   & 0.06  &  2.09  & 2.2  & 684.5 & 3.6 &  5553 &  5547 & 1.09 &        18.0 &      818. \\
13                 Tordera   &0.08  &  2.04  & 2.3  & 688.8 & 3.4 &  7980 &  7977 & 0.76 & 13.6 &     568. \\
14                Baix Ter   & 0.07  &  2.71  & 2.3  & 710.2 & 3.6 &  6042 &  6036 & 1.03 &     17.4 &     746. \\
15            Cap de Creus   &   0.07  &  2.92  & 2.3  & 741.5 & 3.7 &  5962 &  5955 & 1.09 &       17.7 &     754. \\
16    Alt      Llobregat     &  3.12  &  5.82  & 2.6  & 742.8 & 3.3 &  6970 &  6988 & 0.90 &       16.7 &      621. \\
17                    Muga  &   0.06  &  2.56  & 2.4  & 749.3 & 3.6 &  6462 &  6457 & 1.02 &        16.9 &     698.  \\
18             Alt Ter Sau  & 0.08  &  2.43  & 2.5  & 772.1 & 3.6 &  6966 &  6961 & 0.97 &       16.3 &   647. \\
19                Fluvi\`a  &  3.09  &  4.74  & 2.3  & 772.4 & 3.8 &  6287 &  6319 & 1.05  &    16.7 &   697. \\
20         Alt Ter S. Joan  & 0.07  &  1.98  & 2.8  & 795.1 & 3.3 & 8333 & 8327  & 0.84&   15.5 & 452.   \\
\bottomhline
\end{tabular}
\end{table*}

In all the stations, rain is measured
by the same weighing precipitation gauge,
the device called {\it Pluvio} 
from OTT ({\tt http://www.ott-hydrometry.de}),
either with a capacity of 250 or 1000 mm
and working through the balance principle.
It measures both liquid or/and solid precipitation.
The precipitation rate is recorded in intervals of $\Delta t =5$ min, 
with a resolution of $1.2$ mm/hr
(which corresponds to 0.1 mm in 5 min).
This precipitation rate can be converted into an energy flux
through the latent heat of condensation of water, 
%%(2500 kJ/kg at 0$^\circ$C), 
which yields 1 mm/hr $\simeq$ 690 W/m$^2$,
%\sout{
%but we have preferred to work with the
%more easy-to-interpret precipitation rate.
%(notice then that the solar constant puts a limit
%in mean precipitation rate worldwide).
%} %\AD{\hilight{change. ok jefa...}}
 nevertheless, we have not performed such conversion.
 \Fref{Muga_rate} shows a subset of the time series for site 17 (Muga).

 In order to make the files more manageable, the database reports
zero-rain rates only every hour; then we 
  consider time voids larger than 
1 hour
% \sout{have to be considered} 
as operational errors.
The ratio of these missing times to the total time covered
in the record is denoted as $f_M$ in Table \ref{tableone},
where it can be seen that this 
%\sout{ratio} 
is usually below 0.1 \%.
However, there are 3 cases in which its value is
around 3 or 4 \%.
Other quantities reported in the table are
the fraction of time corresponding to rain (or wet fraction), $f_r$,
the annual mean rate, and the mean rate conditioned
to rain periods.
%\AD{Lo puedes explicar mas?}
Nevertheless, note that for a fractal point process a quantity as $f_r$
depends on the time resolution, so, $f_r$
only makes sense for a concrete time division,
in our case, $\Delta t=5$ min.
%\begin{figure}[t]
%\vspace*{2mm}
%\begin{center}
%\includegraphics*[width=6cm%, height=8.5cm
%]{./figs/mapaconquescomarquesestacions.pdf}
%\includegraphics[width=8.3cm,height=4cm]{./figs/mapamediterraneo.pdf}
%\end{center}
%\caption{Map showing the position of the rain gauges analyzed across Catalonia.
%% De donde sale????
%The location of the region in the NW of the Mediterranean is also shown.
%(Coast lines and old political borders taken from {\tt http://rimmer.ngdc.noaa.gov/coast}.)
%%\AD{hacer mapa cuando se pueda}  
%\flabel{map}}
%\end{figure}

\begin{figure}[ht] 
\centering
\subfloat[][]{\flabel{Muga_rate}%
  \includegraphics[width=8cm,angle=0]{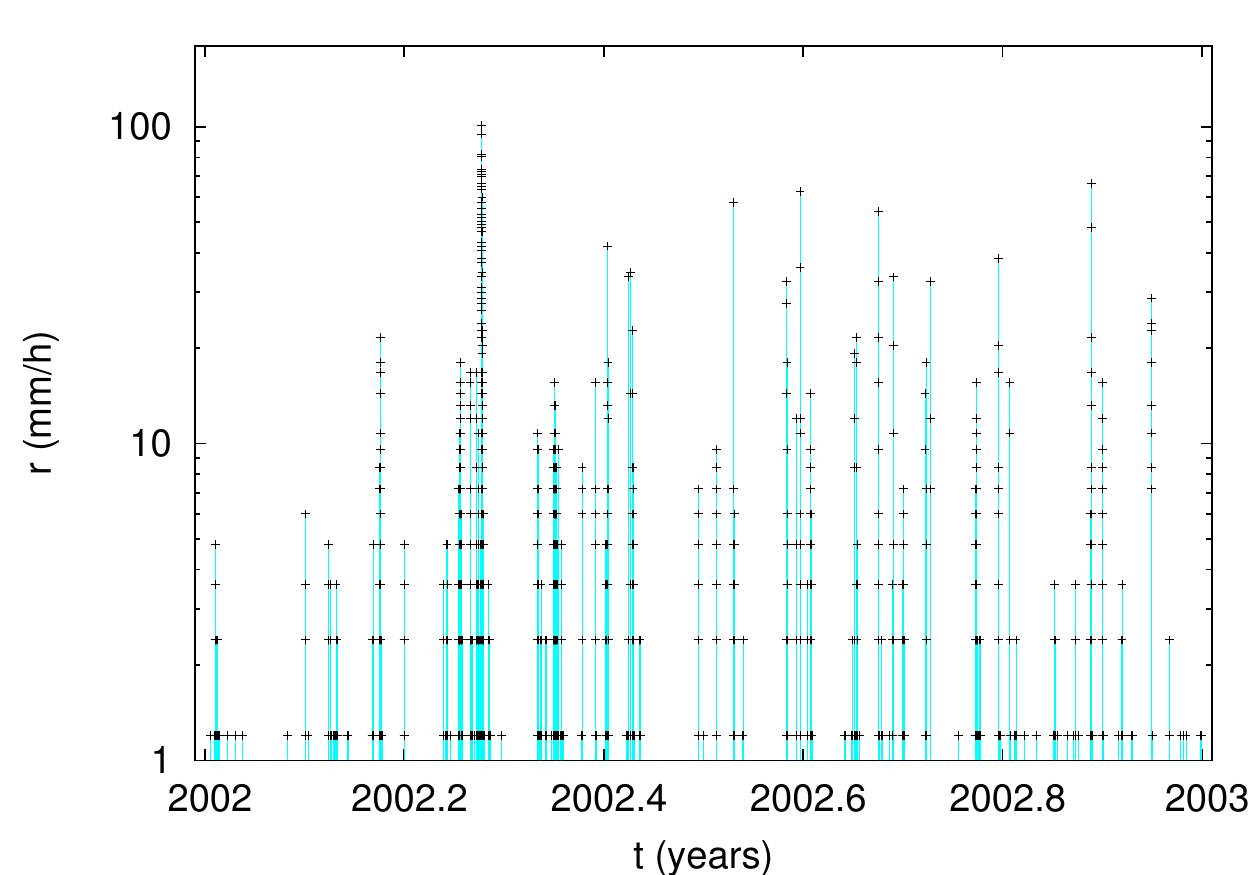}}\\ %}\ %
\vspace{-0.3cm}
\subfloat[][]{\flabel{lluviamasgrande}%
  \includegraphics[height=7.8cm, angle=270]{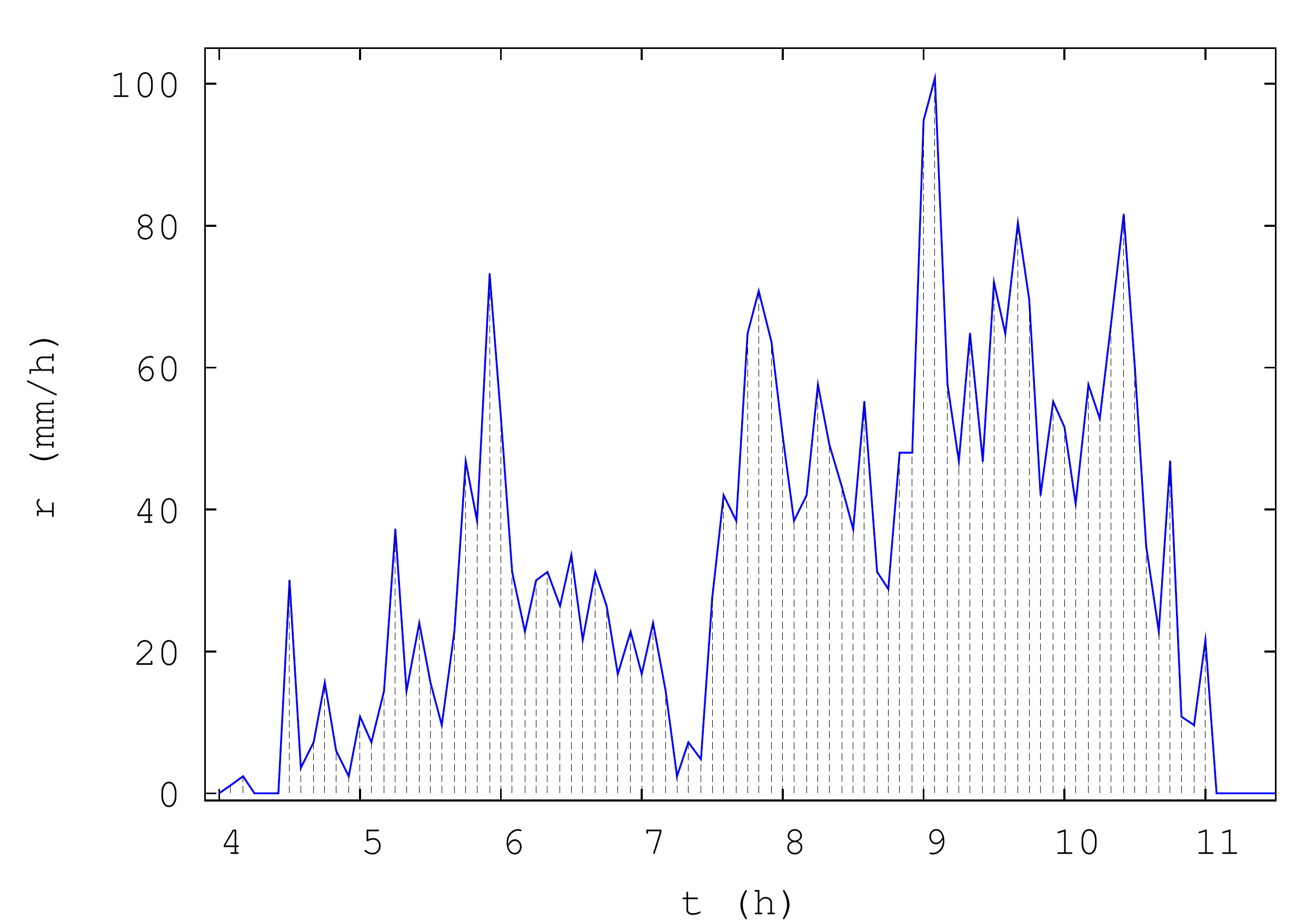}}\\ %}\ %
\vspace{-0.3cm}
\subfloat[][]{\flabel{eventydur}%
  \includegraphics[width=8cm,angle=0]{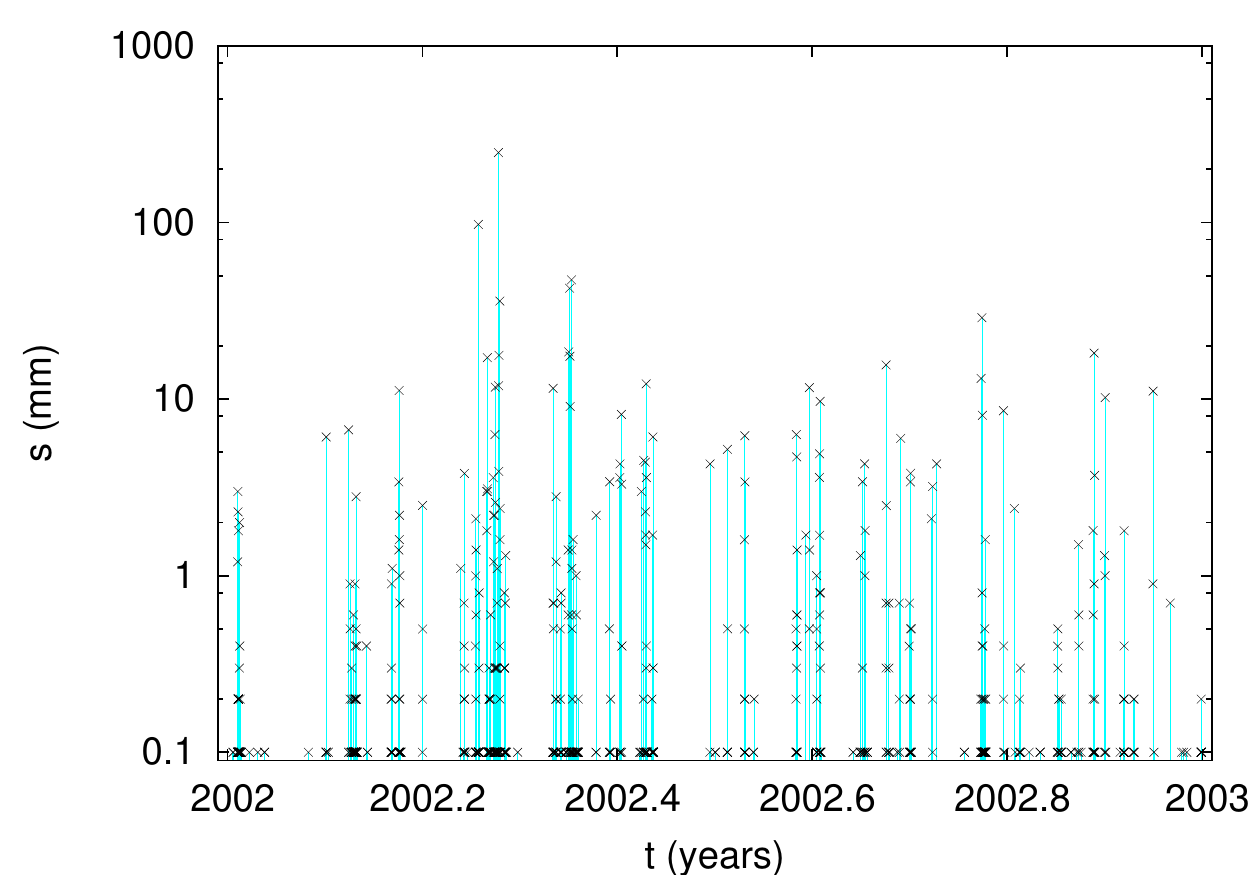}}\ %
\caption{ (a) Subset of the rain rate time series for site 17 (Muga) for year 2002.
 (b)   More reduced subset of the rain rate time series for  the same site, 
 corresponding to the largest rain event on the record, with $s=248.7$ mm, on April 11, 2002.
%{\sout{corresponding to a consecutive non-zero rates (event) with the biggest total rain collected during it (size). 
%This event took place on April 11, 2002, and it is also the largest in size of all sites, with $s=248.7$ mm.}}
Time refers to hours since midnight.
A very small rain event is also present at the beginning, with $s=0.3$ mm and 
separated to the main event by a dry spell of duration $q=15$ min.
(c)  Corresponding event-size time series for the same site, for year 2002. %\sout{Size of all rain events versus their occurrence time in the Muga site.}
\flabel{Muga_rate_and_lluviamasgrande}}
\end{figure}

%%\section{Analysis and Results}

\subsection{%\sout{Rain events and dry spells} 
Rain event sizes, rain event durations, and dry spell durations}

Following \cite{Andrade} and \cite{Peters_prl}, 
we define a rain event as a sequence of consecutive rain rates bigger than zero 
delimited by zero rates, i.e., 
%a certain threshold $c$, i.e.,
$\{r(t_n),r(t_{n+1}),\dots r(t_m)\}$,
such that $r(t_i) > 0 $ for $i=n, n+1, \dots m$
with $r(t_{n-1})=r(t_{m+1})=0$.
Due to the resolution of the record, this
is equivalent to take a threshold with a value below 1.2 mm/hour.
%%(where the superscript means that we are just below the value 1.2).
 It is worth mentioning that this simple definition of rain events may be in conflict
with those used by the hydrologists' community, so caution
is required in order to make comparisons between the different approaches
\citep{Molini}.

 The first 
%\sout{important} 
observable to consider is

%\AD{\hilight{change... vaaale...}}
the duration $d$ of the event, which is the time that the event lasts (a multiple of 5 min, in our case).
The size of the event is defined as the total rain during the event, i.e., the rate integrated over the event 
 duration,
$$
s\equiv \sum_{i=n}^{m} r(t_i) \Delta t \simeq \int_{t_n}^{t_m} r(t) dt,
$$
measured in mm (and multiple of 0.1 mm in our case, 1.2 mm/hr $\times$ 5 min). 
 Notice that this 
 event size 
is not the same as the usual rain depth, 
 due to the different definition of the rain event in each case.
\Fref{lluviamasgrande} shows as an illustration 
the evolution of the rate for the largest event in the record,
which happens at the Muga site,
whereas \fref{eventydur} %(a)
 displays the sequence of all event sizes in the same site
for the year 2002. %The\Fref{eventydur}(b) shows 
 It is important to realize that this quantity is different
to the one at Fig. 1a.
 Regarding event durations, the time series have a certain resemblance to Fig. 1c, 
as usually they are (nonlinearly) correlated with event sizes
%The size of all events in Muga as a function of
%their duration (not shown) has a considerable resemblance to 
\citep{Telesca_rain}.
Further, the dry spells are the periods between consecutive
rain events (then, they verify $r(t)= 0$);
we denote their durations by $q$. 
When a rain event, or a dry spell, is interrupted due to missing data, 
we discard that event or dry spell, and count the recorded
duration as discarded time;
the fraction of these times in the record appears in 
Table \ref{tableone}, under the symbol $f_D$.
{Although in some cases the duration of the interrupted event or dry spell
can be bounded from below or from above (as in censored data), we have not attempted 
 to use that partial information.}

%\end{comment}
%\begin{comment}
\section{Power-law Distributions}

\subsection{%\sout{Rain-event and dry-spell}
  Probability densities}

Due to the enormous variability of the 3 quantities just defined, 
the most informative approach is to work with their probability distributions.
Taking the size as an example, its probability density $P(s)$
is defined as the probability that the size is between $s$ and $s+ds$
divided by $ds$, with $ds \rightarrow 0$. 
Then, $\int_0^\infty P(s) ds = 1$.
This implicitly assumes that $s$ is considered as a continuous variable
(but this will be corrected later,
see more details on Appendix \ref{EstimationDetails}).
%For event durations $d$ and dry spell durations $q$,
%the probability densities are defined in the same way, 
%and denoted as $P(d)$ and $P(q)$,
%with the implicit understanding that their funcional
%forms may be different, despite the use 
%of the same symbol.
In general, we illustrate all quantities with the event size $s$,
the analogous for $d$ and $q$ are obtained by replacing $s$
with the symbol of each observable.  The 
corresponding probability densities are denoted as $P(d)$ and $P(q)$,
with the implicit understanding that their functional
forms may be different. %\AD{\hilight{inconsistent ... yasta...\hilightred{no lo entiendo... a ver asi}}} \hilightgreen{7}.
Note that
the annual number densities \citep{Peters_prl,Peters_pre,Peters_jh}
are trivially recovered multiplying the probability densities
by the total number of events or dry spells and dividing by total time.

The results for the probability densities $P(s)$, $P(d)$, and $P(q)$ 
of all the sites under study 
are shown in 
%{\bf 
Figs. \ref{fig:distributions_event},  \ref{fig:distributions_eventduration}, and \ref{fig:distributions_eventdroughts}, respectively. %}
%%, \fref{distributions_duration} and \fref{distributions_droughts}.
In all cases the distributions show a very clear behavior, monotonically
decreasing and covering a broad range of values.
%(}Remember that a power law appears as a straight line in a double logarithmic plot,
%i.e., $\logP(s) =-\tau_s \log s + \mbox{constant}$).
However, to the naked eye, a power-law range is only apparent
for the distributions of dry spells, $P(q)$
 (remember that a power law turns into a straight line in a log-log plot).
Moreover,  the $P(q)$ are the broadest distributions, covering a range
of more than 4 orders of magnitude (from 5 min to about a couple of months),
and present in some cases 
% and for
%some sites present 
a modest daily peak (in comparison to \cite{Peters_prl},
with 1 day $=$ 1440 min).
In the opposite side we find the distributions of durations, $P(d)$,
whose range is the shortest, from 5 min to about 1 day (two and a half  orders of magnitude), 
and for which no straight line is visible in the plot;
rather, the distributions appear as convex.
The size distributions, $P(s)$, defined
for about 3 orders of magnitude (from 0.1 to 200 mm roughly),
can be considered in between the other two cases, %\sout{ with perhaps}
with a visually shorter range
% \sout{and perhaps with a short range} 
of power-law behavior.

\begin{figure}[ht] 
\centering
\subfloat[][]{\flabel{distributions_event}%
\includegraphics*[width=8cm,angle=0]{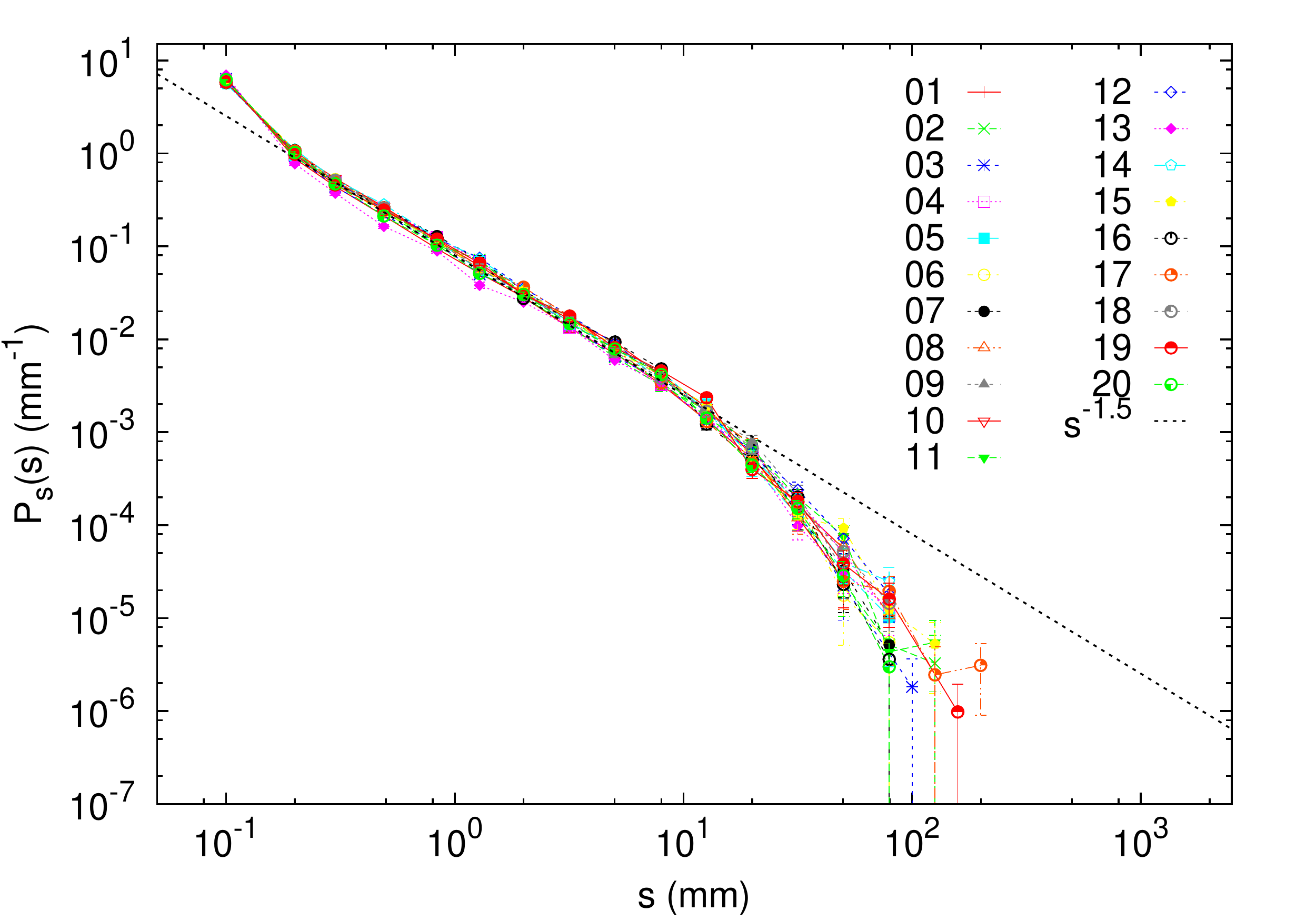}}\\ %}\ %
\vspace{-0.3cm}
\subfloat[][]{\flabel{distributions_eventduration}%
\includegraphics*[width=8cm,angle=0]{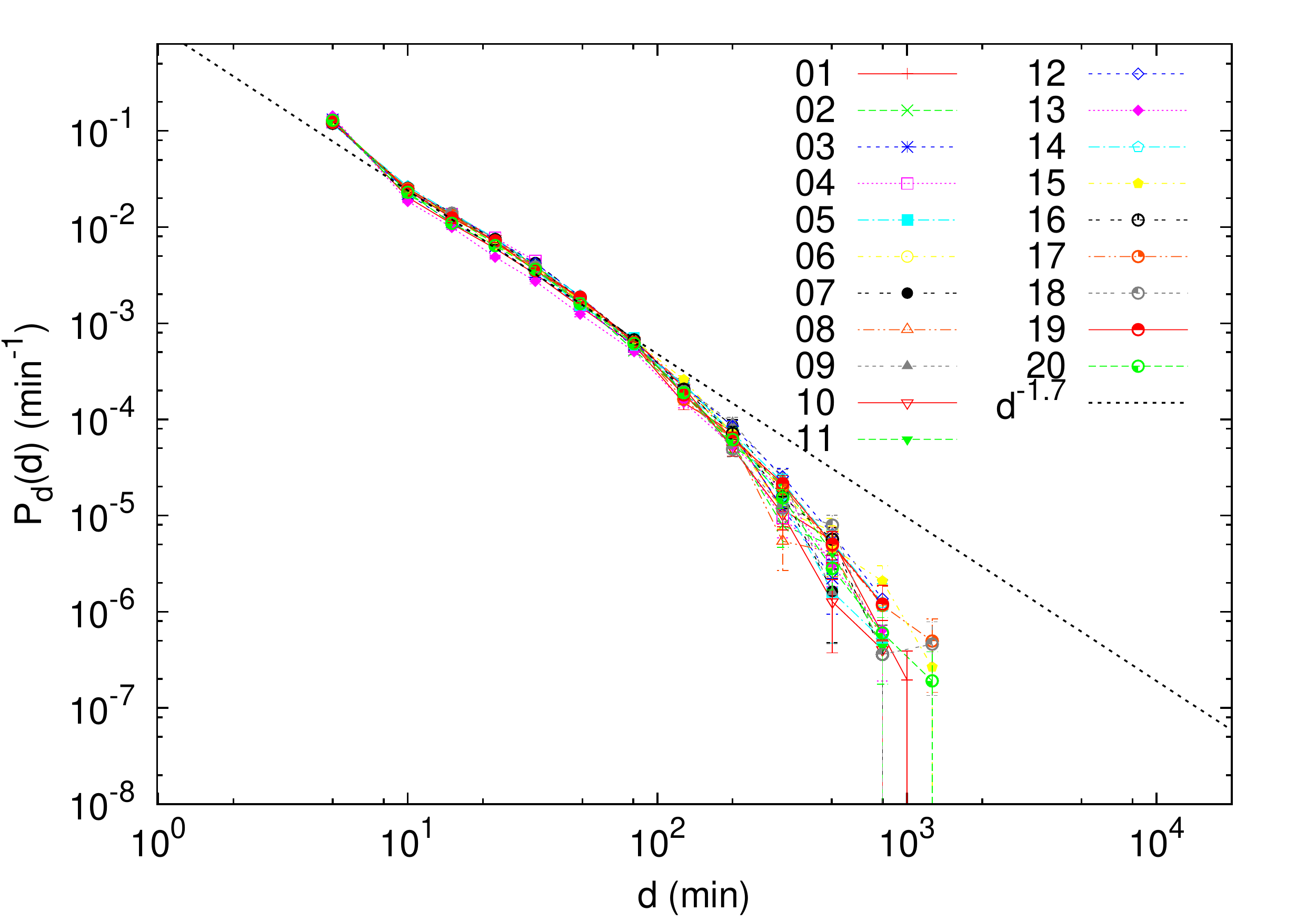}}\\ %}\ %
\vspace{-0.3cm}
\subfloat[][]{\flabel{distributions_eventdroughts}%
\includegraphics*[width=8cm,angle=0]{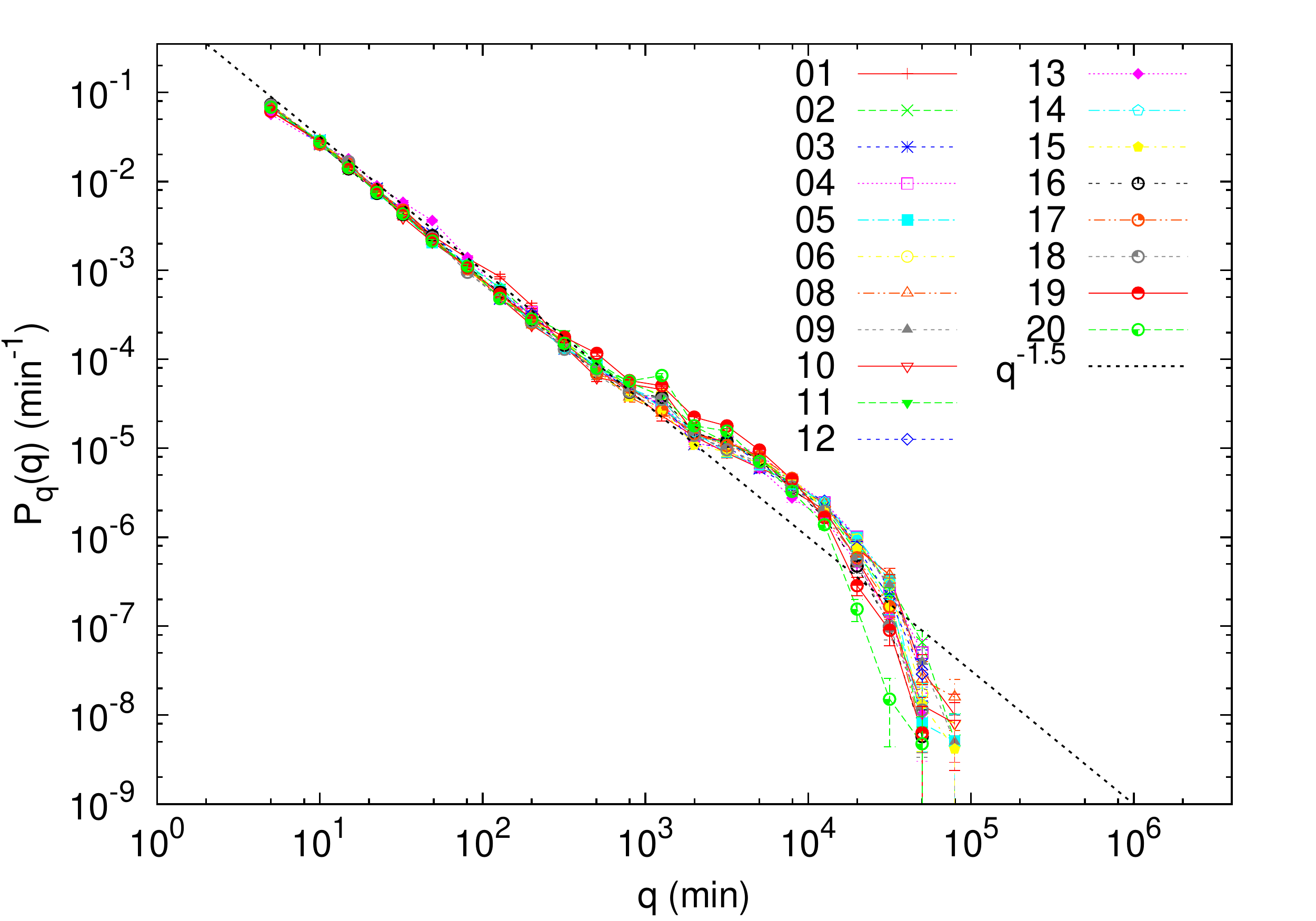}}\\ %}\ %
\caption{   Probability densities for all the sites
for the whole time covered by the record of:
(a)  Event sizes, 
(b) Event durations,
 and 
(c)  Dry spells.
\flabel{distributions}}
\end{figure}

\subsection{Fitting and testing power laws}
A quantitative method can put more rigor into these observations.
The idea is based on the recipe proposed by 
\cite{Clauset} -- see also \cite{Corral_nuclear} -- 
but improved and generalized to our problem.
Essentially, an objective procedure is required
in order to find the optimum range in which a power law
may hold.
Taking again the event size for illustration, 
we report the power-law exponent fit between the values of $\smin$
and $\smax$ which yield the maximum number of data
in that range but with a $p-$value greater than 
$10 \%$.
The method is described in detail in \cite{Peters_Deluca},
but we summarize it in the next paragraphs.

For a given value of the pair $\smin$ and $\smax$,
the maximum-likelihood (ML) power-law exponent is estimated
for the events whose size lies in that range.
This exponent yields a fit of the distribution,
and the goodness of such a fit is evaluated by means of the 
Kolmogorov-Smirnov (KS) test \citep{Press}.
The purpose is to get a $p-$value, 
which is the probability that
% if the data were
%indeed a power law with that exponent, 
the KS test gives a 
distance between 
true power-law data and its fit
larger than 
the distance obtained between 
the empirical data and its fit.
%
%distance between 
%a true empirical power-law distribution 
%(with the ML value of the exponent and the corresponding number of data)
%and a theoretical power-law distribution with the same exponent,
%worse than the one we have obtained with our data.

For instance, $p=20 \%$ would mean that truly
power-law distributed data were closer than the empirical data to their respective fits 
in $80 \%$ of the cases,
but in the rest $20 \%$ of the cases a true power law were at a larger distance
than the empirical data.
So, in such a case the KS distance turns out to be somewhat large, 
but not large enough to reject that 
%the ML power law is a good fit to the data.
the data follow a power law with the ML exponent.

As in the case in which some parameter is estimated from the data
there is no closed formula to calculate the $p-$value, 
we perform Monte Carlo simulations in order 
to compute the statistics of the Kolmogorov-Smirnov distance
and from there the $p-$value.
In this way, for each $\smin$ and $\smax$
we get a number of data $\bar N_s$ in that range and,
 repeating the procedure many times, a $p-$value.
We look for the values of the extremes ($\smin$ and $\smax$) which maximize
the number of data in between but with the restriction that
the $p-$value has to be greater than $10 \%$
(this threshold is arbitrary, but the conclusions
do not change if it is moved).
The maximization is performed sweeping 100 values of $\smin$ and 100 values of $\smax$,
in log-scale,
in such a way that all possible ranges (within this log-resolution) are taken into account.
%{\bf COMO QUEDA ESTO???
We have to remark that, in contrast with \cite{Peters_Deluca},
we have considered always discrete probability distributions, both in the ML fit
and in the simulations.
%} \hilightgreen{8 Bueno, se puede quedar asi porque no especificamos el metodo...
%al menos lo hemos intentado en discreto!!}
Of course, it is a matter of discussion which approach
(continuous or discrete) is more appropriate
for discrete data that represent a continuous process.
In any case, the differences in the outcomes are rather small.
%%\AD{ seguro??}
Notice also that the method is not based on the estimation of the probability densities
shown in the previous subsections,  what would be inherently more arbitrary
\citep{Clauset}.

The results of this method are in agreement with the visual conclusions
obtained in the previous subsection, as can be seen in Table \ref{results}.
Starting with the size statistics, 13 out of the 20 sites yield 
reasonable results, with an exponent $\tau_s$ between 1.43 and 1.54
over a logarithmic range $\smax/\smin$ from 12 to more than 200.
For the rest of the sites, the range is too short, less than one decade
%{\bf
 (a decade is understood from now as an order of magnitude).
In the application of the algorithm, it has been necessary
to restrict the value of $\smin$ to be $\smin \ge 0.2 $ mm;
otherwise, as
%and the exponent too large (sometimes larger than 2).
%The reason is that 
the distributions have a concave shape (in logscale)
close to the origin
(which means that there are many more events 
in that scale than at larger scales),
%Then, 
the algorithm (which maximizes the number of data in a given range)
prefers a short range with many data close to the origin
than a larger range with less data away from the origin.
%We conclude that the results we obtain are due
%to the inadequacy of the algorithm rather than to 
%an abscense of power-law behavior in the data.
{It is possible that a variation of the algorithm in which 
the quantity that is maximized  were different
(for instance related with the range),
would not need the restriction  in the 
%\sout{sizes} 
 minimum size. %\AD{\hilight{SOLO SIZES?}} 
%\hilightgreen{9 ok}

For the distribution of durations 
%%the results are worse, 
%%as expected 
%%\sout{(worse in the sense that the power-law fit is worse
%%than in the previous case).} {\bf :}
the resulting power laws turn out to be
very limited in range;
only 4 sites %\sout{do not give totally unacceptable} 
 give not too short power laws, %\sout{ranges,} 
with 
$d_{max}/d_{min}$ from 6 to 12
and 
$\tau_d$ from 1.66 to 1.74.
The other sites yield 
 extremely short ranges 
for the power law be of any relevance.
%%and too large exponents, in general.
The situation is analogous to the case of the distribution
of sizes, 
but the resulting ranges are much shorter here \citep{Peters_Deluca}.
%in which the algorithm looks for the 
%maximum number of data which is obtained 
%in the short range close to the origin.
Notice that the excess of events with $d=5$ min,
eliminated from the fits imposing $d_{min} \ge 10$ min,
has no counterpart in the value of the smallest rate
(not shown), and therefore,  we conclude that this extra number of events 
is 
%\sout{probably} 
due to problems in the time resolution of the data. %\AD{\hilight{Ccambiar. yasta}}
%We have had to impose $d_{min} \ge 10$ min
%in order to avoid this problem.

Considerably %\sout{better} 
more satisfactory are the results for the dry spells.
16 sites give consistent results, 
with $\tau_q$ from 1.45 to 1.55
in a range $q_{max}/q_{min}$ from 30 to almost 300.
It is noticeable that in these cases $q_{max}$ is always below 1 day.
 %\sout{Perhaps} 
The removal by hand of dry spells around that value
should enlarge a little the power-law range. %\AD{\hilight{cambiar. yasta.}}
In the rest of sites, either the range is comparatively too short
(for example, for the Gai\`a site, the power-law behavior of $P(q)$
is interrupted at around $q=100$ min), or the algorithm
has a tendency to include the bump the distributions show
between the daily peak ($q$ beyond 1000 min)
and the tail.
This makes the value of the exponent smaller
 (around 1.25). Nevertheless, the value of the exponent is much higher than the
one obtained for the equivalent problem of 
earthquake waiting times, 
where the Omori law leads to values around one, 
or \
%sout{smaller} 
less.
%{\bf
 This points to a fundamental differences%\hilight{\sout{s}}
between both kind of processes (from a statistical point of view).
%}
%% \cite{Corral_Christensen}.}

In summary, the power laws for the distributions of durations are too
short to be relevant, and the fits for the sizes are in the limit
of what is acceptable  (some cases are clear and some other not).
%%Most of this failure is probably a problem with the algorithm.
Only the distributions of dry spells give really good power laws,
with $\tau_q=1.50 \pm 0.05$, and for more than two decades in 6 sites.

%\begin{widetext}
\begin{table*}[t]
 \caption{ 
Results of the power-law fitting and goodness-of-fit tests
applied to event sizes, event durations, and dry-spell durations (in mm or in min),
for the period of 9 and a half years specified in the main text.
The table displays the minimum of the fitting range, $s_{min}$, 
and the ratio between the maximum and the minimum of the fitting range (logarithmic range, $s_{max}/s_{min}$),
 %and maximum fitting range, 
%%the ratio of these values (logarithmic range), 
total number of events, number of events in fitting range
($\bar N_s$, $\bar N_d$, and $\bar N_q$, for $s$, $d$, and $q$, respectively),
and the power-law exponent with its uncertainty (one standard deviation)
calculated as stated by \cite{Bauke}
and displayed between parenthesis as the variation of the last digit.
\label{results}}
%\hspace{-1.65cm}
\centering
\begin{tabular}{r r r r r r}
%%\hline % la quito porque si no no se ve la barra
%\textbf
{Site}& $s_{min}$  & $\frac{s_{max}}{s_{min}}$ & $N_s$ & $\bar{N_s}$  & $\tau_s$ \\ 
        %&         mm   &      &          &          &                         \\
\hline %\hline
%tiny{
 1      &         0.2 &       180.5 &     5393 &     1886 &    1.54(2)  \\
 2       &         0.2 &      4.5 &     5236 &     1323 &    1.64(6)  \\
 3       &         0.2 &        155.5 &     5749 &     2111 &    1.53(2)  \\
 4       &         0.2 &        12.0 &     5108 &     1745 &    1.43(3)  \\
 5      &         0.2 &          140.0 &     5289 &     2106 &    1.52(2)  \\
 6       &         0.2 &         68.0 &     4924 &     1969 &    1.49(2)  \\
 7      &         0.2 &        105.5 &     5219 &     2234 &    1.51(2)  \\
 8       &         0.2 &          213.0 &     5112 &     2047 &    1.53(2)  \\
 9       &         0.2 &           5.0 &     5366 &     1459 &    1.53(5)  \\
10       &         0.2 &         19.0 &     6691 &     2452 &    1.51(2)  \\
11       &         0.2 &            65.0 &     6224 &     2373 &    1.49(2)  \\
12       &         0.2 &           4.0 &     5967 &     1500 &    1.53(6)  \\
13       &         0.3 &            66.7 &     8330 &     1853 &    1.45(2)  \\
14       &         0.2 &          3.5 &     6525 &     1711 &    1.56(6)  \\
15       &         0.3 &            3.7 &     6485 &     1102 &    1.39(7)  \\
16       &         0.2 &           3.5 &     7491 &     1852 &    1.59(5)  \\
17       &         0.2 &            80.5 &     6962 &     2853 &    1.52(2)  \\
18       &         0.2 &          41.5 &     7511 &     2847 &    1.51(2)  \\
19       &         0.2 &           99.5 &     6767 &     2742 &    1.47(2)  \\
20       &         0.2 &             3.5 &     9012 &     2047 &    1.69(5)  \\
%\hline
%PCh      
  %       &         0.01&         300 &  -- &     -- &    1.4  \\
\hline
\end{tabular}
%  \caption{ Nmaxpval10 integer
%   \label{table: Nmaxpval10}}
%\end{table}
%\end{widetext}
%\begin{widetext}
%\begin{table}[h]
%\centering
\begin{tabular}{ r r r r r r|} %\hline
 $d_{min}$    & $\frac{d_{max}} {d_{min}}$  & $\bar{N_d}$  & $\tau_d $ \\ 
    %   min      &                             &              &                                \\
\hline %\hline
             10 &      10.0 &          1668 &    1.67(4)   \\
             10 &          4.0 &          1581 &    1.60(6)   \\
             10 &           6.0 &          1726 &    1.66(5)   \\
             10 &             3.5 & 1564 &    1.41(7)   \\
             10 &              3.5 &          1530 &    1.58(7)   \\
             10 &              3.5 &          1441 &    1.59(7)   \\
             10 &            3.5 &          1621 &    1.51(7)   \\
             10 &             4.0 &          1567 &    1.55(6)   \\
             10 &           4.0 &          1658 &    1.51(6)   \\
             10 &            3.5 &          2066 &    1.57(6)   \\
             10 &            5.0 &          1932 &    1.56(5)   \\
             10 &           4.0 &          1889 &    1.49(6)   \\
             10 &          12.5 &          2288 &    1.74(3)   \\
             10 &           5.0 &          2299 &    1.62(4)   \\
             10 &           5.0 &          2095 &    1.64(5)   \\
             10 &             4.0 &          2385 &    1.59(5)   \\
             10 &           3.5 &          2087 &    1.60(6)   \\
             10 &             3.5 &          2238 &    1.57(6)   \\
             10 &              3.5 &          1958 &    1.60(6)   \\
             10 &               8.5 &          2972 &    1.66(3)   \\
 \hline
 %            10 &            30  &           --  &   1.6    \\
 %\hline
\end{tabular}
%  \caption{ Nmaxpval10 integer
%   \label{table: Nmaxpval10}}
%\end{table}
%\end{widetext}
%
%\begin{widetext}
%\begin{table}[h]
%\centering
\begin{tabular}{ r r r r r r }%\hline
 $q_{min}$  & $\frac{q_{max}}{q_{min}}$ & $N_q$ & $\bar{N_q}$  & $\tau_q $ \\ 
      %  min     &      &          &          &                         \\
\hline %\hline
       95 &          7.8 &     5387 &      743 &    1.75(7) \\
        5 &         273.0 &     5231 &     4729 &    1.46(1) \\
       10 &          80.0 &     5745 &     3207 &    1.53(2) \\
        5 &           196.0 &     5103 &     4520 &    1.47(1) \\
       20 &            47.3 &     5287 &     1706 &    1.45(2) \\
       20 &           31.3 &     4926 &     1537 &    1.47(3) \\
        5 &         256.0 &     5215 &     4734 &    1.51(1) \\
       15 &         65.0 &     5107 &     2098 &    1.50(2) \\
       10 &          90.0 &     5419 &     2889 &    1.55(2) \\
       25 &         33.0 &     6685 &     1758 &    1.48(3) \\
       45 &       468.3 &     6219 &     2005 &    1.24(1) \\
        5 &          235.0 &     5961 &     5376 &    1.47(1) \\
      130 &         158.5 &     8328 &     1501 &    1.27(2) \\
        5 &         215.0 &     6520 &     5906 &    1.47(1) \\
       15 &         49.7 &     6479 &     2560 &    1.51(2) \\
        5 &       214.0 &     7510 &     6789 &    1.50(1) \\
       10 &      68.5 &     6958 &     3719 &    1.52(1) \\
       20 &             31.0 &     7507 &     2302 &    1.53(2) \\
       50 &       21.7 &     6800 &     1378 &    1.26(3) \\
       15 &          34.3 &     9007 &     3367 &    1.50(2) \\
\hline
 %      5   &     $^*$1200 &     --    &     --  &    1.4 \\\hline
\end{tabular}
\\
\hfill
%$^*$ Disregarding the daily peak.
\end{table*}
%\end{widetext}

%\AD{parametric non-parametric: What to change to satisfy Referee 2:
%- He\/ she does not understand sec 3.3 and sec 3.4 titles)
%
%- Complains about the algebra, he does not understand the interest
%of changing of variables, clarify.

%- Complains about no study the dependence on the sample size of the 
%so-called finite-size effects.

%- Not clear for him the collapse on figures 5,6

%- large uncertanties on fig7 

%}
%\begin{comment}
\section{Scaling}

\subsection{Non-parametric scaling}

 However, the fact that a power-law behavior does not exist
over a broad range of values does not rule out the existence
of SOC \citep{Christensen_Moloney}.
In fact, the fulfillment of a power-law distribution in the form of Eq. (\ref{powerlaw})
is only valid when finite-size effects are ``small'',
which only happens for large enough systems.
In general, when these effects are taken into account, 
SOC behavior leads to distributions of the form \citep{Christensen_Moloney,Peters_Deluca},
\begin{equation}
P(s)=s^{-\tau_s} \G_s(s/s_\xi) %\mathrm{ for } 
 \mbox{ \, for  \,} s > s_l,
\label{density}
\end{equation}
where
%$\tau_s > 1$ for normalization in the limit $s_\xi \rightarrow \infty$,
%and 
$\G_s(x)$ is a scaling function that is
essentially constant  for $x \ll 1$ and decays fast for $x \gg 1$,
accounting in this way for the finite-size effects 
when $s$ is above the crossover value $s_\xi$;
the size $s_l$ is just a lower cutoff limiting the validity
of this description.
The pure power law only emerges for $s_\xi \rightarrow \infty$,
nevertheless, a truncated power law holds 
%%%Notice that for a power-law behavior to hold 
over an appreciable range
%%%it is necessary that
if
the scales given by $s_l$ and $s_\xi$ are well separated,
i.e., $s_l \ll s_\xi$. 
As $s_\xi$ increases with system size, typically as $s_\xi \propto L^{D_s}$
(with $D_s$ the so-called avalanche dimension, or event-size dimension), 
the power-law condition (\ref{powerlaw})
can only be fulfilled for large enough system sizes.

 Note that, in the case of a too short power-law range or a non-conclusive fit, 
we still could check the existence of scaling using Eq. (\ref{density})
if we knew $s_\xi$ or $L$.
However, $s_\xi$ is difficult to measure, needing a parameterization of the scaling function,
and it is not clear what the system size $L$ is for rainfall.
 It could be the vertical extension of the clouds, 
or the depth of the troposphere.
Nevertheless, it is important to realize that the scaling ansatz (\ref{density}) still can be checked from data 
 without knowledge of $L$ or $s_\xi$.
First, notice that the ansatz implies that the $k-$order moment of $s$
scales with $L$ as 
\begin{equation}
\langle s^k \rangle \propto L^{D_s(k+1-\tau_s)}
% \mathrm{ for } 
 \mbox{ \, for  \,} 1 < \tau_s < k+1,
%% also tau > 1??
\label{moments}
\end{equation}
if $s_l \ll s_\xi$, see \cite{Christensen_Moloney}.
Second, Eq. (\ref{density}) can be written in a slightly different form,
as a scaling law, 
\begin{equation}
P(s)=L^{-D_s \tau_s} \mathcal{F}_s(s/L^{D_s}) 
%\mathrm{ for } 
 \mbox{ \, for  \,} s > s_l,
\label{density_bis}
\end{equation}
where the new scaling function $\mathcal{F}_s(x)$ is defined as
$\mathcal{F}_s(x) \equiv  x^{-\tau_s} \G_s(x/a)$
($a$ is the constant of proportionality between $s_\xi$ and $L^{D_s}$).
This form of $P(s)$ (in fact, $P(s,L)$), with an arbitrary $\mathcal{F}$, 
is the well-known scale-invariance condition 
 for functions with two variables \citep{Christensen_Moloney}.
Changes of scale (linear transformations) in $s$ and $L$ may leave
the shape of the function $P(s,L)$ unchanged
 (this is what scale invariance really means,
power laws are just a particular case in one dimension).

Substituting $L^{D_s} \propto \ave{s^2}/\ave{s}$
and $L^{D_s\tau_s}\propto L^{2 D_s}/\ave{s} \propto \ave{s^2}^2/\ave{s}^3$
(from the scaling of $\ave{s^k}$, assuming $\tau_s < 2$)
into Eq. (\ref{density_bis}) leads to
\begin{equation}
P(s)= \ave{s}^3 \ave{s^2}^{-2}
\tilde{\mathcal{F}}_s(s \ave{s} / \ave{s^2}),
% \mathrm{ for } s > s_l,
\label{density_tris}
\end{equation}
where $\tilde{\mathcal{F}}_s(x)$ is essentially the scaling function 
${\mathcal{F}}_s(x)$, absorbing the proportionality constants.
Therefore, if scaling holds, a plot of $\ave{s^2}^2 P(s)/\ave{s}^3 $ 
versus $s \ave{s} / \ave{s^2}$ for all the sites has to yield
a collapse of the distributions into a single curve, 
which draws $\tilde{\mathcal{F}}_s(x)$
(a similar procedure is outlined in \cite{Rosso}).
In order to proceed,
the mean and the quadratic mean, $\ave{s}$ and $\ave{s^2}$, can be easily estimated
from data.
 Since no estimation of parameters is involved for this procedure, 
we call it non-parametric scaling.
%These values, and the corresponding ones for $d$ and $q$ are displayed for all sites in Table \ref{tableone}.

The outcome for $P(s)$, $P(d)$, and $P(q)$ is shown in 
%%\fref{distributions_event_rosso}, \fref{distributions_eventduration_rosso} and \fref{distributions_eventdroughts_rosso},
 Figs. \ref{fig:distributions_event_rosso}, \ref{fig:distributions_eventduration_rosso}, and \ref{fig:distributions_eventdroughts_rosso},
with reasonable results, especially
for the distribution of dry spells. 
The plot suggests that the scaling function $\G_q$
of the dry-spell distribution
has a maximum around $x\simeq 1$,
 but this does not in disagreement with our approach,
which only assumed a constant scaling function for small $x$
and a fast decay for large $x$.

Note that the quotient $\ave{s^2}/\ave{s}$ gives the scale for
the crossover value $s_\xi$ (as $s_\xi \propto \ave{s^2}/\ave{s}$,
with a constant of proportionality that depends on the scaling function $\G_s$
and on $s_l/s_\xi$), 
and therefore it is the ratio of the second moment to the mean
and not the mean which describes the
scaling behavior of the distribution.
 This can have important implications for extreme events:
an increase in the value of the mean is not proportional to an increase
of the most extreme events, represented by $s_\xi$.
For the case of event sizes, we get values of $\ave{s^2}/\ave{s}$ between 10 and 30 mm
%%(see Table \ref{tableone}),
(which is a variability much higher than that of $\ave{s}$),
and therefore the condition $s_l \ll s_\xi$
is very well fulfilled (assuming that the moment ratio $\ave{s^2}/\ave{s}$ 
is of the same order as $s_\xi$, and with $s_l \simeq s_{min}$),
which is a test for the consistency of our approach.
For dry spells $\ave{q^2}/\ave{q}$ is between 5 and 13 days,
which is even better for the applicability of the scaling analysis.
The case of the event durations is somewhat ``critical'',
with $\ave{d^2}/\ave{d}$ between 70 and 120 min,
which yields $d_\xi/d_l$ in the range from 14 to 24.
Nevertheless, we observe that the condition $s_l \ll s_\xi$
for the power law to show up is stronger than the same
condition for the scaling analysis to be valid.

%larger that the 3 days found for the dry-spell crossover in Ref. \cite{Peters_pre}
%associated to the time scale given by a passing frontal 
%weather system
%(nevertheless, 
%we do not know how the crossover was defined in that paper
%and the comparison is not really possible).

%%%%%%%%%%%%%%%%%%%%%%% rescale rosso%%%%%%%%%%%%%
\begin{figure}[ht] 
\centering
\subfloat[][]{\flabel{distributions_event_rosso}%
\includegraphics*[width=8cm,angle=0]{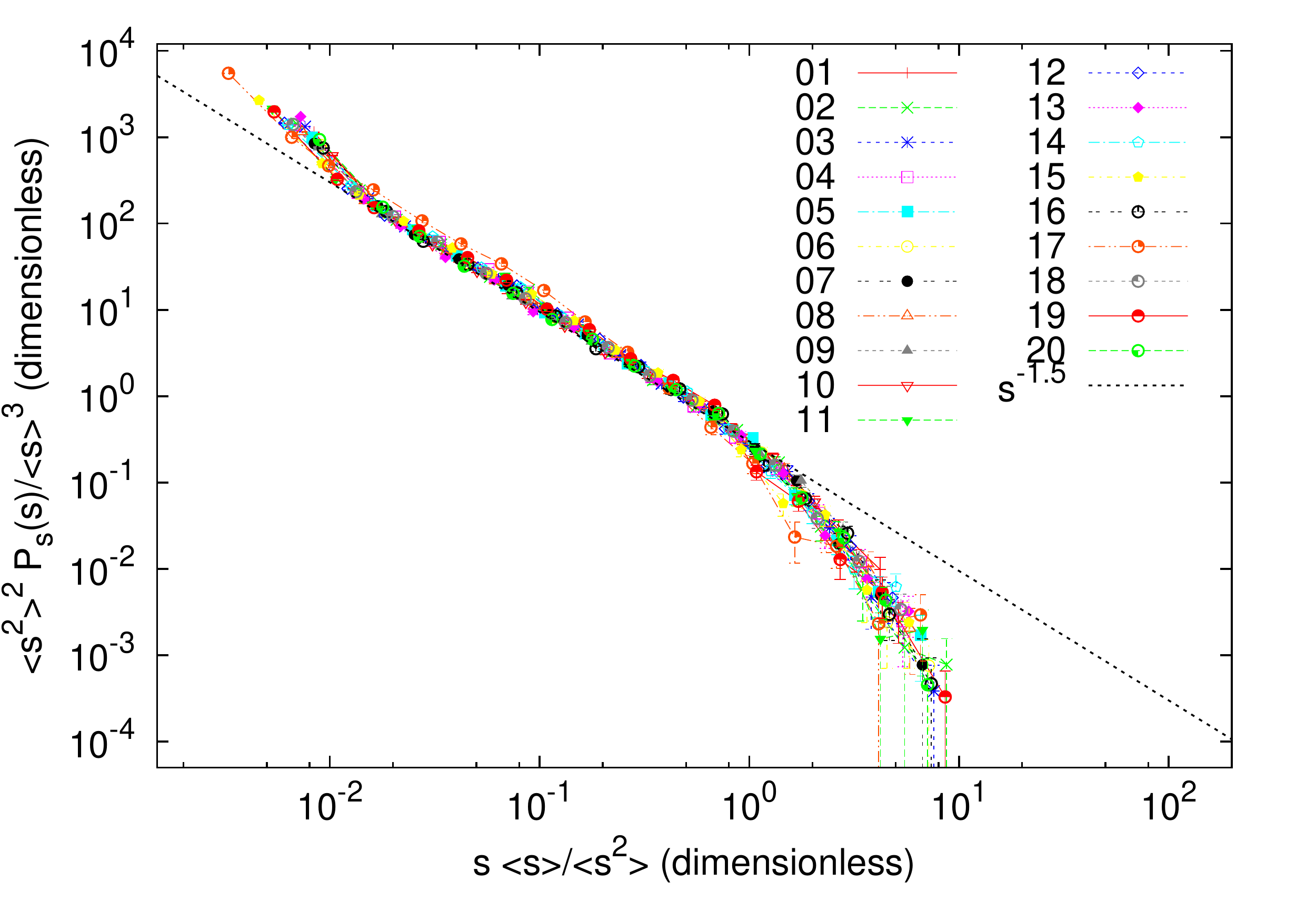}}\\ %}\ %
\vspace{-0.3cm}
\subfloat[][]{\flabel{distributions_eventduration_rosso}%
  \includegraphics*[width=8cm,angle=0]{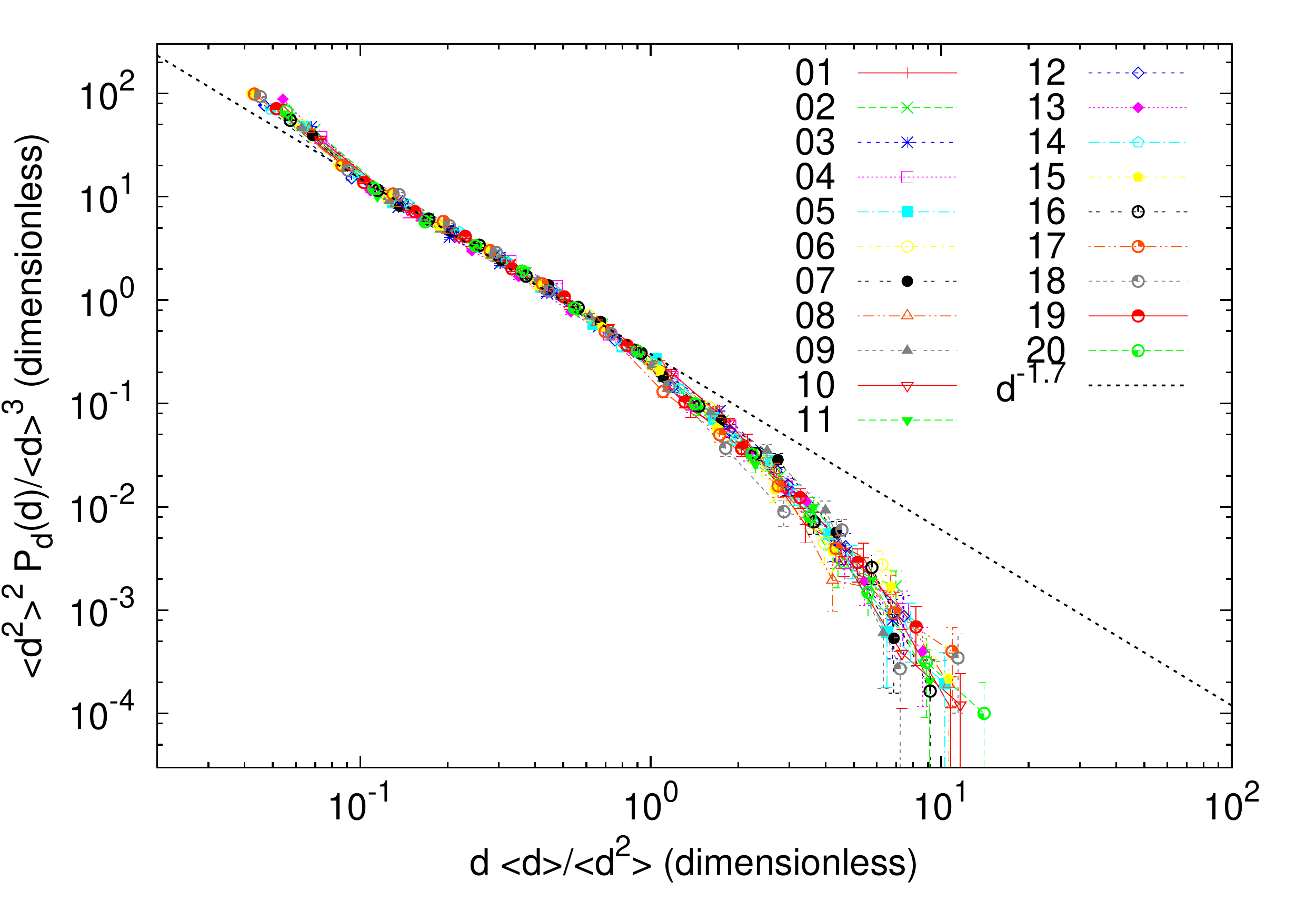}}\\ %}\ %
\vspace{-0.3cm}
\subfloat[][]{\flabel{distributions_eventdroughts_rosso}%
  \includegraphics*[width=8cm,angle=0]{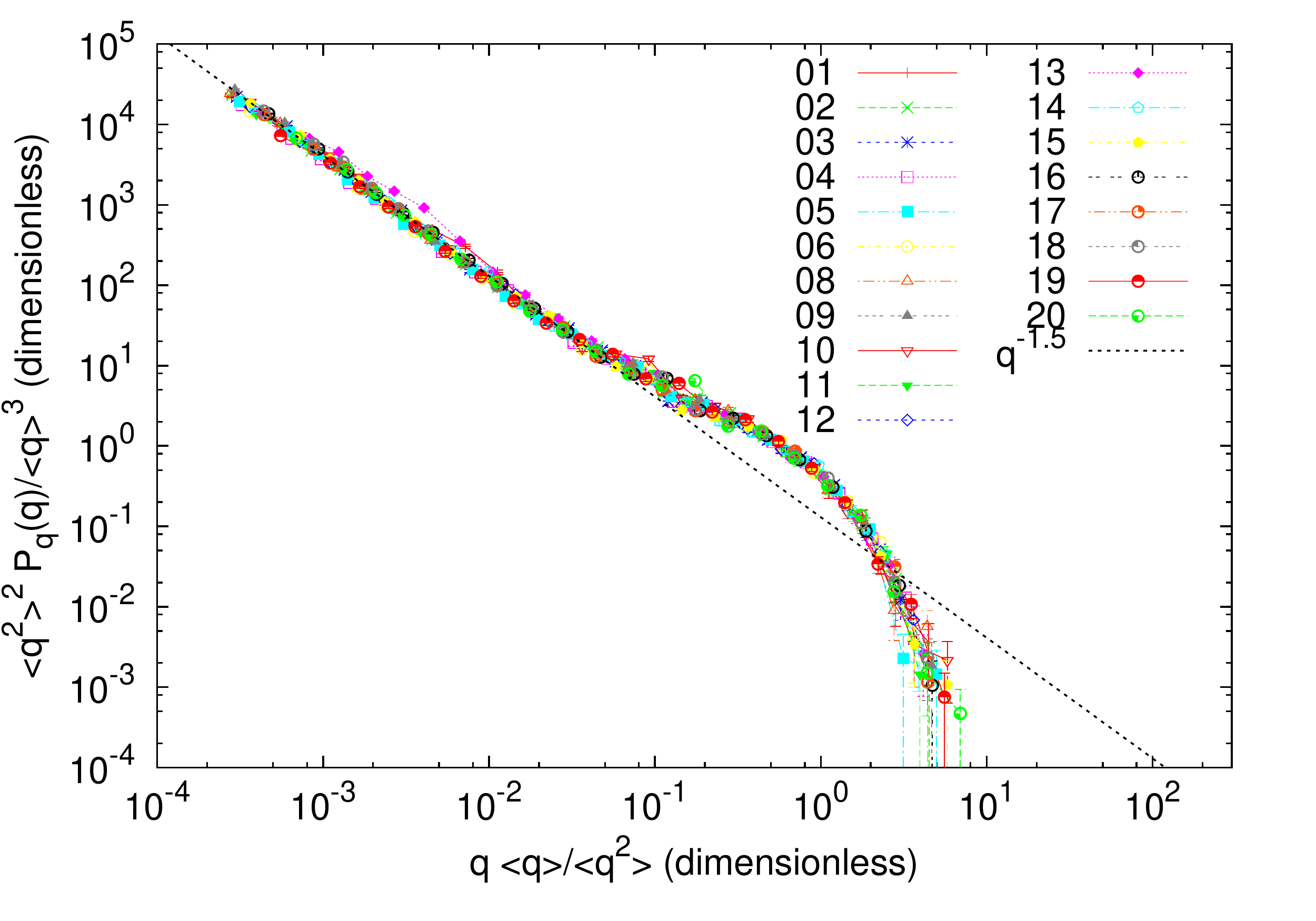}}\ %
\caption{ { 
Collapse of the probability densities for all the sites
for the whole time covered by the record of:
(a) Event sizes,  
(b) Event durations, 
and 
(c) Dry spells.  Rescaling is performed} using the first and second moment 
 of each distribution, following Eq. (\ref{density_tris}).
\flabel{distributions_rosso}}
\end{figure}

\subsection{Parametric scaling}
\label{sub: Parametric scaling}

Further, a scaling ansatz as Eq. (\ref{density}) or (\ref{density_bis}) 
allows an estimation
of the exponent $\tau_s$, even in the case in which a power law
cannot be fit to the data.
From the scaling of the moments of $s$ we get, taking $k=1$,
$L^{D_s} \propto \ave{s}^{1/(2-\tau_s)}$ and
$L^{D_s\tau_s} \propto \ave{s}^{\tau_s/(2-\tau_s)}$
(again with $\tau_s < 2$); 
so, substituting into Eq. (\ref{density_bis}),
\begin{equation}
P(s)= \ave{s}^{-\tau_s/(2-\tau_s)} \hat{\mathcal{F}}_s(s/\ave{s}^{1/(2-\tau_s)}).
% \mathrm{ for } s > s_l,
\label{density_cuat}
\end{equation}
One only needs to find the value of $\tau_s$ that optimizes the collapse
of all the distributions, i.e., that makes the previous equation valid, 
or at least as close to validity as possible.
 As the scaling depends on the parameter $\tau_s$,
we refer to this procedure as parametric scaling.

We therefore need a measurement to quantify distance between rescaled distributions.
In order to do that, we have chosen to work with the cumulative distribution function,
$S(s) \equiv \int_{s}^{\infty} P(s')ds'$, rather than with the density
(to be rigorous, $S(s)$ is the complementary of the cumulative distribution function,
and is called survivor function or reliability function in some contexts).
 Although in practice both $P(s)$ and $S(s)$ contain the same probabilistic information,
the reason to work with $S(s)$ is double: the estimation of the cumulative distribution function does
not depend of an arbitrarily selected bin width $ds$ \citep{Press},
and it does not give equal weight to all scales
in the representation of the function (i.e., in the number of points that constitute the function).
The scaling laws (\ref{density_bis}) and (\ref{density_cuat}) turn out to be, then,
\begin{equation}
S(s)=L^{-D_s (\tau_s-1)} \mathcal{H}_s(s/L^{D_s}),
%\mathrm{ for } s > s_l,
\end{equation}
\begin{equation}
S(s)=\ave{s}^{-(\tau_s-1)/(2-\tau_s)} \hat{\mathcal{H}}_s(s/\ave{s}^{1/(2-\tau_s)}),
% \mathrm{ for } s > s_l,
\end{equation}
with ${\mathcal{H}_s}(x)$ and $\hat{\mathcal{H}_s}(x)$
the corresponding scaling functions.

The first step of the method of collapse
is to merge all the pairs $\{s,S(s)\}_i$
into a unique rescaled function $\{x,y\}$.
If $i=1,\dots, 20$ runs for all sites, 
and $j=1,\dots, M_s(i)$ for all the different values that the size of events takes 
on site $i$ (note that $M_s(i) \le N_s(i)$),
then,
$$ %\begin{equation}
x_\ell(\tau) \equiv \log (s_{ji}/\ave{s}_i^{1/(2-\tau)}),
$$
$$
%%y_\ell(\tau) \equiv \log (S_{s\, i}(s_{ji}) \ave{s}_i^{\tau/(2-\tau)}),
y_\ell(\tau) \equiv \log (S_{i}(s_{ji}) \ave{s}_i^{(\tau-1)/(2-\tau)}), % compruebalo hija
$$ %% \end{equation}
with $s_{ji}$ the $j$-th value of the size in site $i$,
$\ave{s}_i$ the mean on $s$ in $i$,
$S_{i}(s_{ji})$ the cumulative distribution function in $i$,
and $\tau$ a possible value of the exponent $\tau_s$.
The index $\ell$ labels the new function, from 1 to $\sum_{\forall i} M_s(i)$,
in such a way that $x_\ell(\tau) \le x_{\ell+1}(\tau)$;
i.e., the pairs $x_\ell(\tau), y_\ell(\tau)$ are sorted by increasing $x$.

Then, we just compute
\begin{equation}
D(\tau) \equiv \sum_{\forall \ell}
\left(
[x_\ell(\tau) - x_{\ell+1}(\tau)]^2 +
[y_\ell(\tau) - y_{\ell+1}(\tau)]^2
\right),
\end{equation}
which represents the sum of all Euclidean distances
between the neighboring points in a (tentative) collapse plot in logarithmic scale.
The value of $\tau$ which minimizes this function is
identified with the exponent $\tau_s$ in Eq. (\ref{density}).
We have tested the algorithm applying it to SOC models 
whose exponents are well known (not shown).

The results of this method applied to our datasets, not only for the size distributions
but also to the distributions of $d$, %%and $q$, 
are highly satisfactory.
There is only one requirement:
the removal of the first point in each distribution
($s=0.1$ mm and $d=5$ min), 
as with the ML fits.
%the removal of the first point in the event-duration distributions
%($d=5$ min), 
%as with the ML fits.
The exponents we find are 
%$\tau_s=1.22$, $\tau_s=1.30$, and $\tau_s=1.23$,
$\tau_s=1.52 \pm 0.12$ and $\tau_d=1.69 \pm 0.01$,
%\AC{error demasiado pequenyo!!!!??? como puede ser???} \AD{\hilightred{INVESTIGANDO}}
% \hilightgreen{10}
in agreement with the ones obtained by the power-law fitting 
method presented above;
%\sout{and} 
the corresponding rescaled plots are shown in \fref{distributions_collapse}.
 Although the visual display does not allow to evaluate properly the quality
of the collapse, the reduction in the value of the function $D(\tau)$ is notable. Then,
the performance of the method is noteworthy, taking into account that
the mean values of the distributions show little variation in most cases.
In addition, the shape of the scaling function $\G_s$
can be obtained by plotting, as suggested by Eq. (\ref{density}),
$s^{\tau_s}P(s)$ versus $s/\ave{s}^{1/(2-\tau_s)}$,
and the same for the other variable, $d$. %% and $q$.
%\sout{
 \Fref{flat} displays what is obtained for each distribution.
%}. 
%{\bf Por que esta tachado esto??
%No nos referimos a la figura nunca mas!!!}
 In contrast, the application of this method to $P(q)$
does not yield consistent results, as $\tau_q$ turns out to be 
rather small (1.24).
Notice that the existence of a daily peak in the distributions 
is an obstacle to a data collapse, as the peak prevents 
a good scaling.

\begin{figure}[ht] 
\centering
\subfloat[][]{\flabel{distributions_event_collapse}%
\includegraphics*[width=8cm,angle=0]{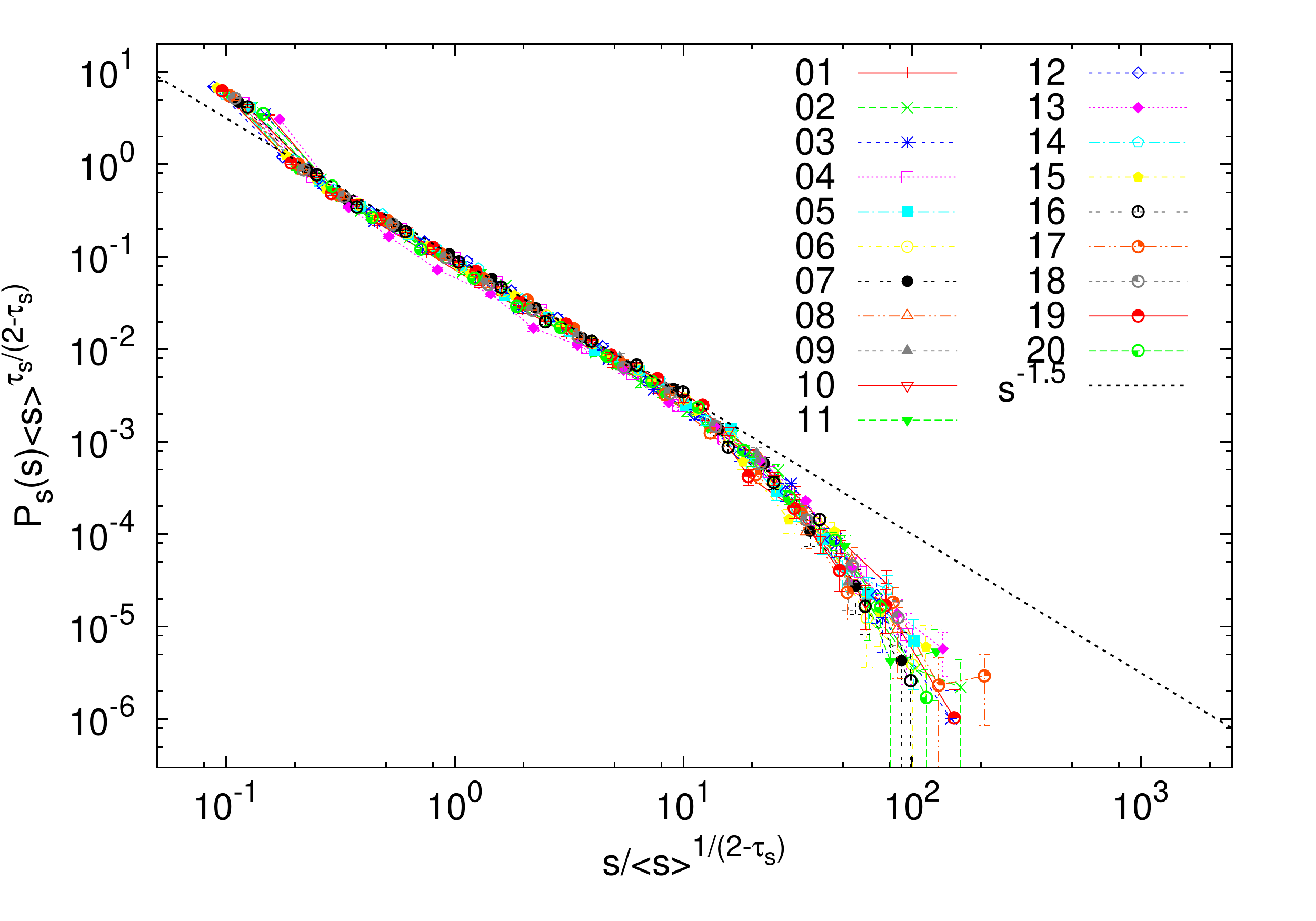}}\\ %}\ %
\vspace{-0.3cm}
\subfloat[][]{\flabel{distributions_eventduration_collapse}%
  \includegraphics*[width=8cm,angle=0]{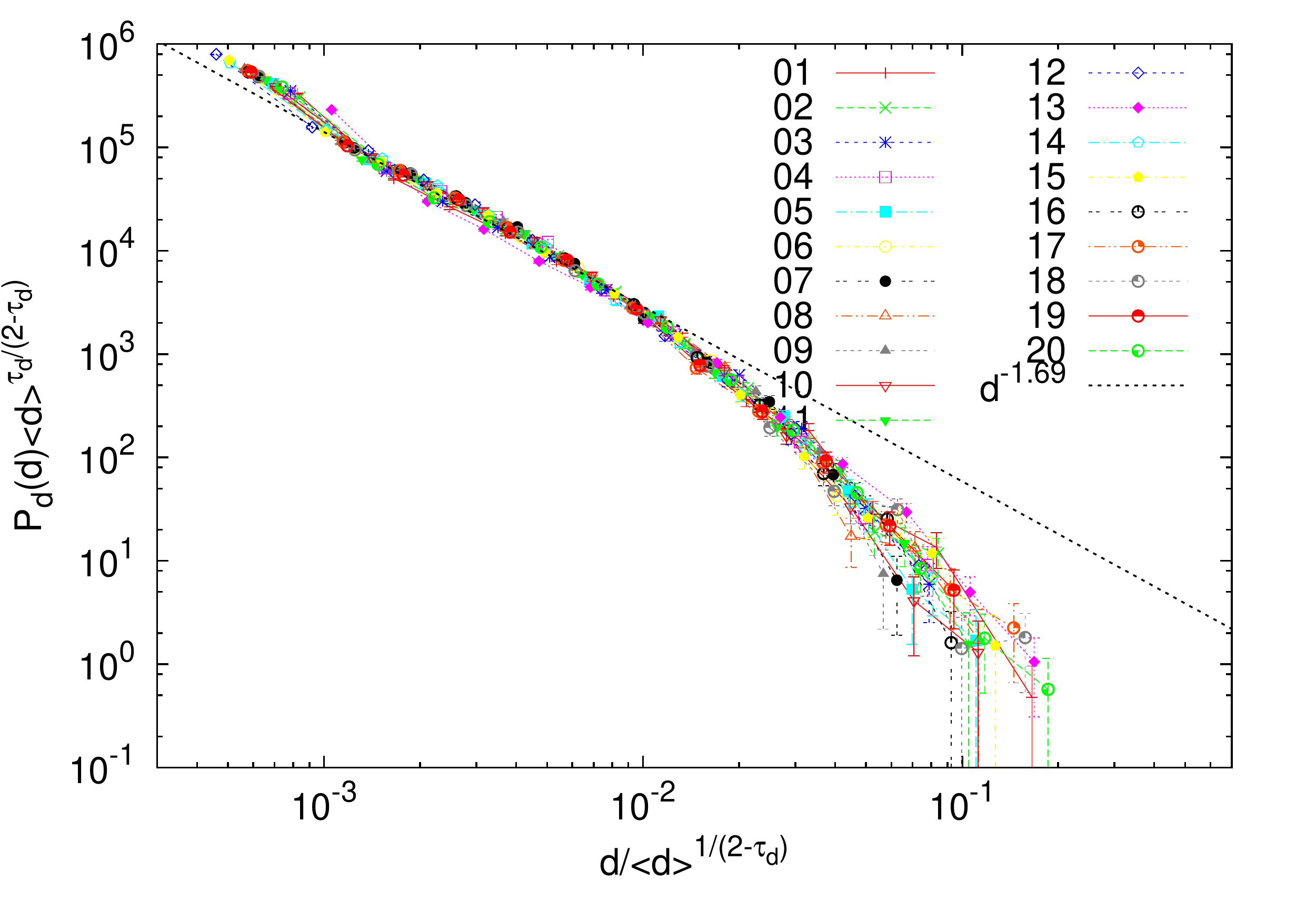}}\\ %}\ %
%\vspace{-0.3cm}
%\subfloat[][]{\flabel{distributions_eventdroughts_collapse}%
%  \includegraphics*[width=8.3cm,angle=0]%{./figs/err_density_c_eventdroughts_aca_collapse_newformat.pdf}}\ %
\caption{ %{\bf
Collapse of the probability densities for all the sites
for the whole time covered by the record of:
(a) Event sizes  and  
(b) Event durations; 
%\sout{, 
%and 
%(c) Dry spells;} 
rescaled using  Eq.  (\ref{density_cuat}) with
 the exponents: $\tau_s=1.52$  and $\tau_d=1.69$, %\sout{and $\tau_q=1.23$} 
determined minimizing the Euclidean distance between parametrically collapsed distributions.
%\sout{see sec. \ref{sub: Parametric scaling}.} 
Units are mm or min to the corresponding powers appearing in the axes. %}
%{Units are mm$^{-(\tau_s-1)/(2-\tau_s)}$ for the abscissa
%and %%\sout{mm$^{\tau_s-1}$}
%mm$^{2(\tau_s-1)/(2-\tau_s)}$
%for the ordinate in (a), 
%and  min$^{-(\tau_d-1)/(2-\tau_d)}$ for the abscissa
%and min$^{2(\tau_d-1)/(2-\tau_d)}$ for the ordinate in (b) and (c).
%% and (c),
%%with $\tau=\tau_d$ and $\tau=\tau_q$, respectively.  
%}
\flabel{distributions_collapse}}
\end{figure}

%\begin{comment}

%%%%%%%%%%%%%%%%%%%%%%% flat %%%%%%%%%%%%%
%\begin{comment}
\begin{figure}[ht]
%\vspace*{2mm}
%\begin{center}
\centering
\subfloat[][]{\flabel{distributions_event_collapse_scafun}%
\includegraphics*[width=8cm, angle=0]{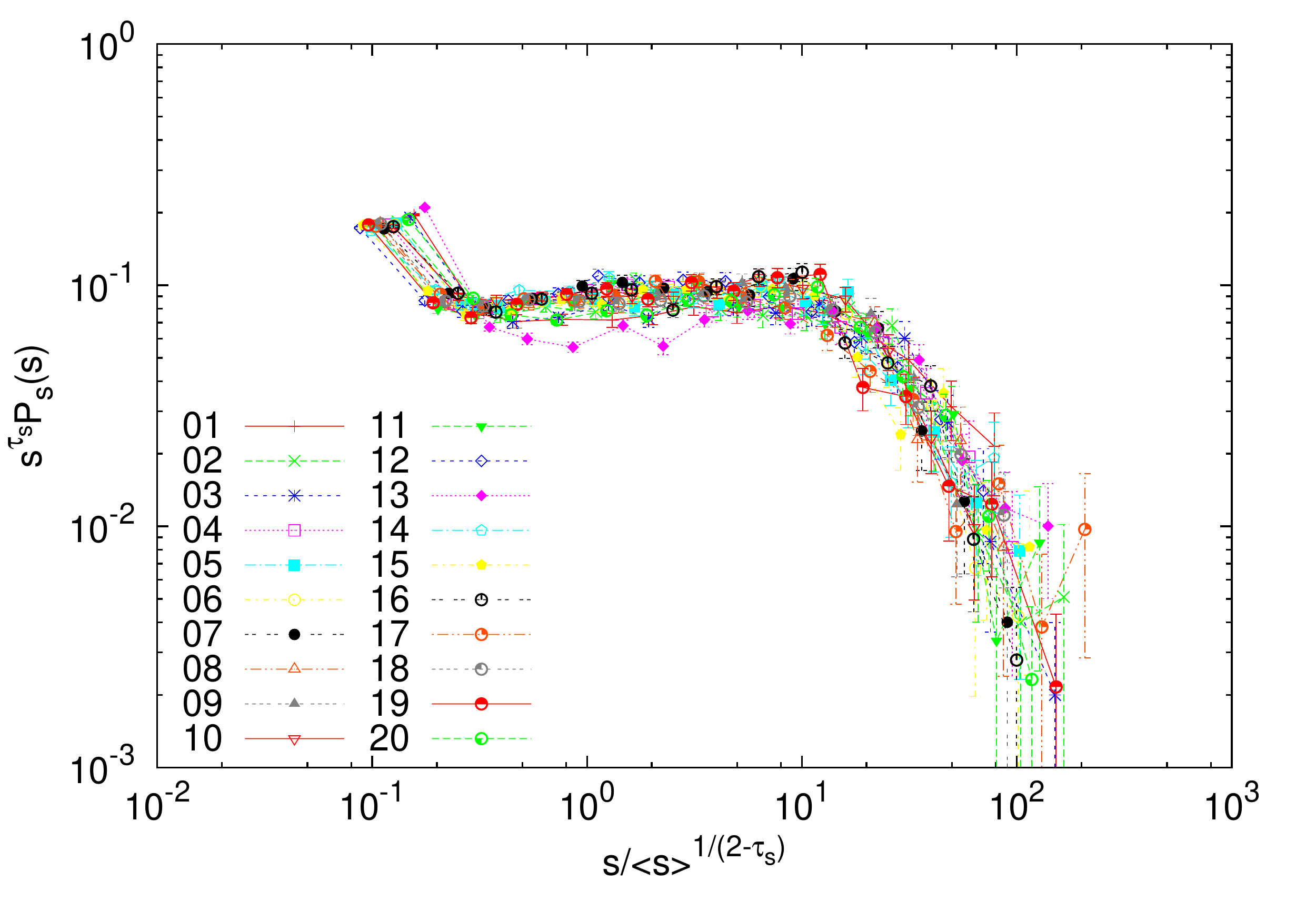}}\\
\vspace{-0.3cm}
\subfloat[][]{\flabel{distributions_eventduration_collapse_scafun}%
\includegraphics*[width=8cm, angle=0]{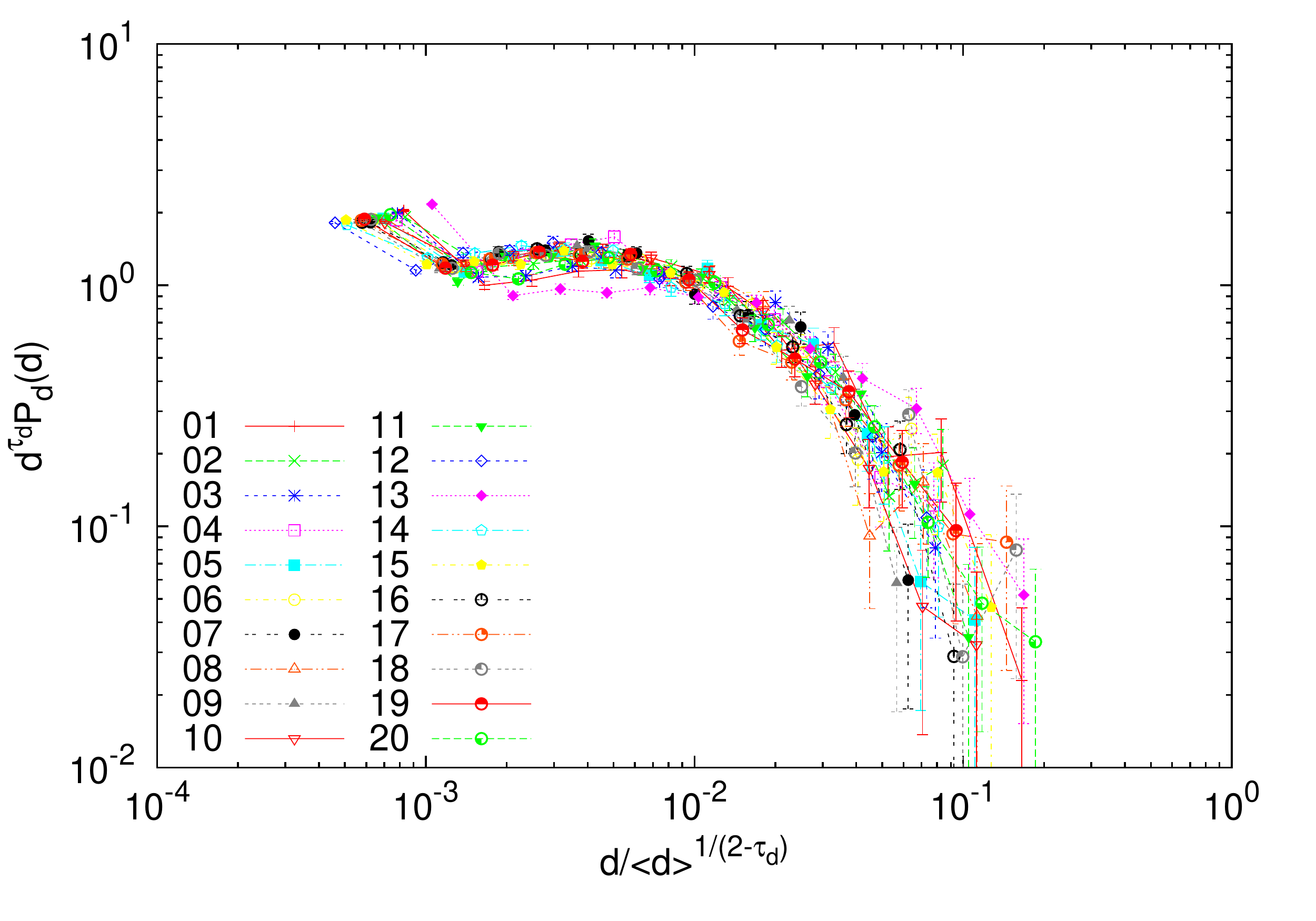}}\\
%\vspace{-0.3cm}
%\subfloat[][]{\flabel{distributions_eventdroughts_collapse_scafun}%
%\includegraphics*[width=8.3cm, angle=0]{./figs/err_density_c_eventdroughts_aca_collapse_scafun_newformat.pdf}}\
\caption{
Inferred scaling functions $\G_s$  and $\G_d$ %\sout{and $\G_q$}
corresponding to the rescaled distributions of  $s$ and $d$ % \sout{, and $q$}}
in \fref{distributions_collapse}, 
multiplied by $s^{\tau_s}$  and $d^{\tau_d}$. %\sout{, or $q^{\tau_q}$} 
Units in the abscissae are as in the previous plot, 
whereas in the ordinates these are mm$^{\tau_s-1}$  and min$^{\tau_d-1}$. %\sout{and min$^{\tau_q-1}$.} 
%\sout{Units are mm or min to the corresponding powers appearing in the axes.}
}
\flabel{flat}
\end{figure}
%\end{comment}

\conclusions[Discussion and Conclusions]
 %\hilightgreen{GENERAL REVISION NEEDED}
%% \conclusions[modified heading if necessary]

%% \conclusions[modified heading if necessary]

%\AD{Referee 2:

%- Conclusions vague... try to be more direct with language 

%we obtained evidence that rainfall in the ... is compatible...
%... it is not essentiallyy different to what is expected from...

%    \hilight{- Highlighted things MUST CHANGE}

%}
%{\bf P1. Summary fit exponent results sizes and durations AND DRY SPELLS!!}

We have performed an in-depth study of the properties of SOC related observables in 
 rainfall in the Mediterranean region
in order to check %\sout{the hypothesis that}
 if this framework can be useful for 
%\sout{model the rain events sizes and durations, and the dry spells durations.}
 modeling rain events and dry spells.
The results support %\sout{to} 
this hypothesis, which had not been checked before in this region or for this kind of data resolution.
%\sout{We have obtained evidence that rainfall in the Mediterranean
%region studied is compatible
% with self-organized criticality.}
For the distributions of rain-event sizes we get power-law exponents 
%\sout{which are} 
valid for one or two decades
% \sout{at least one decade or even two}
 in the majority of sites, with exponent values $\tau_s \simeq 1.50 \pm 0.05$. For the distributions
of event durations, 
the fitting ranges 
are 
%\sout{much} 
shorter,
reaching in the best case one decade, with exponents $\tau_d \simeq 1.70 \pm 0.05$.
This range is expected to be shorter than for event sizes, given that 
%\sout{event durations} 
 these combine the event %\sout{size}
  duration distribution with the rain rate \citep{Peters_Deluca}.
And finally, the dry spell distributions yield the more %\sout{significant} 
 notable
%\sout{remarkable}
 power law fits,
with exponents in the range $\tau_q \simeq 1.50 \pm 0.05$,
in some cases for more than 2 decades.

%{\bf P2. Comparison with data resolution, mention to thresholding effect}

These results are compatible with the ones %{\bf 
obtained for
%}
 the Baltic sea by \cite{Peters_prl},
%{\bf 
which yielded  $\tau_s \simeq\tau_q \simeq 1.4$ and $\tau_d\simeq 1.6$.
%}
%{\bf 
 The agreement is %\sout{notable}
  remarkable,
%}
 taking into account the 
different nature of the data analyzed and the disparate fitting procedures. 
%\sout{where  $\tau_s \simeq\tau_q \simeq 1.4$, and $\tau_d\simeq 1.6$.}
However, %{\bf 
the concordance
%}
 with the 
%{\bf
 more recent
%}
 results of \cite{Peters_Deluca} is %\sout{\sout{not so} {\bf un}clear}
not very good, quantitatively. 
 That previous %\sout{This previous} 
study,
%\sout{W}
with a minimum detection rate of 0.2 mm/hr and a time resolution $\Delta t=1$ min, found
%\sout{it was found there that }%% for several sites across the globe, 
$\tau_s \simeq 1.18$  for several sites 
%\sout{with disparate climatic characteristics}
 across different climates, using essentially the same statistical techniques
as  %\sout{here} 
in the present study. 
Exponents were found not universal for durations %{\bf 
of events and
%} 
dry-spells, 
but 
%{\bf 
for the latter
%}
 they were close in many cases to $\tau_d\simeq1.3$. 
The difference between the 
%\sout{ {\bf event}
 size and dry-spell duration} exponents may be due to 
%\sout{that} 
data resolution. 
Changes in the detection
threshold 
%\sout{has}
 have a non-trivial repercussion in the size and duration of the events
and the dry spells 
(an increase in the threshold can split one single event into two or more
separate ones but also can remove events). Further, 
better time resolution and lower detection threshold 
allow the detection of smaller events, 
enlarging the power law range and reducing the weight from the part close to the crossover point (where the distribution becomes 
steeper); this trivially leads to smaller values of the exponent.
% far from the influence of the crossover point, which leads to a decrease weight of the tail, explaining trivially a smaller exponent value.
%and then due to the presence of the tail
%\sout{we get} 
 %explaining trivially a larger exponent value.
%{\bf LA FRASE ANTERIOR NO SE ENTIENDE!!!}
In the dry spells case, the power-law range in this study is 
enough to guarantee that
our estimation of the exponents are robust, so the discrepancy with
 \cite{Peters_Deluca} may be due to the non-trivial effect of the change in the detection thresholds or differences on the measurement devices.

%{\bf P3. Collapse results}

%{\bf 
 On the other hand, finite size effects can explain the limited power-law range obtained for $s$ and $d$ observables,
%}
%\sout{The limited power-law range obtained for both observables ($s$ and $d$) can be 
%explained by the existence of finite size effects,}
% \sout{it is the case}
 as it occurs in other (self-organized and non-self-organized) critical phenomena.
%\sout{A} 
The finite-size analysis  performed, in terms of combinations of powers of the moments of the distributions,
supports this conclusion. The collapse of the distributions  is a clear 
signature of scale-invariance: different sites share a common shape 
of the rain-event and dry-spell distributions, 
%\sout{and}
  with 
%\sout{the only}
 differences
%\sout{is} 
in the scale of those distributions,
%\sout{which depends} 
depending on 
%{\bf the} 
system size. 
 Then, in the ideal case
%\sout{, in}
 of an infinite system, the power laws  would lack %\sout{of} 
an
 %\sout{could then be extended with no } 
upper cutoff. 
 Moreover, the collapse of the distributions allows %\sout{one} 
an independent 
estimate of the power-law exponents,  which,  for event durations and sizes, 
are in surprising agreement
with the values obtained by the maximum-likelihood fit.
 For dry spell durations, a daily peak in the distributions hinders their collapse.

Nevertheless, future work should consider spatially extended events. 
Our measurements are taken in a point of the system 
which reflects information on the vertical scale, then,
the results could be affected by this. 
%\sout{(Obviously, with poor resolution data the detection 
%threshold cannot be decreased, but it can be artificially increased in very accurate data.)} 
Another important 
 issue are the implications of the results for hazard assessment.
%\sout{ is risk and predictions implication of the results}. % the late studies on risk and prediction implications which . 
% paradigm implications for risk and Finally, note that the presence of SOC in rainfall has 
%important consequences for the risks posed by this natural phenomenon.
If there is not a characteristic rain-event size, then there
is neither a definite separation nor a fundamental difference between the 
smallest harmless rains and the most hazardous storms.
Further, it is generally believed that the critical evolution of events
in SOC systems implies that, at a given instant, it is equally likely that
the event intensifies or weakens, 
%the outcome depending
%on a myriad of uncontrollable details, 
which would make detailed prediction 
%from the time series of events
unattainable. %, in principle.
However, this view has been recently proved wrong,
as it has been reported that a critical evolution describes the dynamics
of some SOC systems only 
on average;
{further, the existence of finite size effects
% \sout{can facilitate} 
can  be used for prediction
\citep{Garber_pre,Martin_Paczuski}. 
%%%%,Garber_pre}.
%%Our findings imply that this could be the case for Mediterranean storms;
Interestingly, in the case of rain, %%nevertheless, 
it has been recently shown by 
\cite{Molini}
 that
knowledge of internal variables of the system
allows some degree of prediction
for the duration of the events,
related also to the departure of the system
from quasi-equilibrium conditions.
%\sout{Therefore, } 
 Finally, we 
%\sout{also} 
urge studies which explore the effects of resolution and detection-threshold 
value in high-resolution rain data. { A common SOC misbelief is that  
avalanches happen following a memoryless process, leading therefore to exponential
distributions for the waiting times \citep{Corral_comment}. 
This has been proved wrong if a threshold on the intensity is present 
\citep{paczuski2005interoccurrence}.
% \AD{Falta cita Maya PRL 2005}.
 In this case,
times between avalanches follow a power-law distribution, as we find for dry spells. %it is found for the  }
%{\bf Our results for the dry spells distributions confirm this.}
%LA FRASE ANTERIOR QUEDA DESCOLGADA DE LAS OTRAS...???
}

In summary, we conclude that the statistics of rainfall events in the NW Mediterranean area studied 
%\sout{is not essentially different to what is expected from}
  are in agreement with the SOC paradigm  expectations.
%\sout{and} 
 This is the first time this study is realized for this region and it is a confirmation 
 %\sout{essentially} 
of what has been found  for other places of the world, 
but using 
%\sout{data with higher resolution there }
 %\sout{using} 
in  our}case data with lower resolution. %\citep{Peters_Deluca}. 
If a representative universal exponent existed, this would
mean that just one parameter is enough for characterizing the distributions. This 
would indicate that the rain event observable cannot detect climatic
differences between regions, but would shed light on universal properties
and mechanisms of rainfall generation.

\appendix

\section{\\ \\Details on the estimation of the probability density}
\label{EstimationDetails}

In practice, the estimation of the density from data is performed
taking a value of $ds$ large enough to guarantee statistical significance,
and then compute $P(s)$ as $n(s)/(N_s \Delta)$,
where $n(s)$ is the number of events with size in the range between $s$ and $s+ ds$,
$N_s$ the total number of events, 
and $\Delta$ is defined as 
%the number of possible different values of the variable 
%in the interval considered; this is given by
$$
\Delta = R_s(\lfloor (s+ds)/R_s \rfloor - \lfloor s/R_s\rfloor),
$$
with
$\lfloor x \rfloor $ the integer part of $x$ and 
$R_s$ the resolution of $s$, i.e. $R_s=0.1$ mm
(but note that high resolution means low $R_s$). 
So, $\Delta/R_s$ is the number of possible different values of the variable 
in the interval considered.
Notice that using $\Delta$ instead of $ds$ in the denominator of the estimation of $P(s)$
allows one to take into account the discreteness of $s$.
If $R_s$ tended to zero, then $\Delta \rightarrow ds$
and the discreteness effects would become irrelevant.

How large does $ds$ have to be to guarantee the statistical significance
of the estimation of $P(s)$?
Working with long-tailed distributions
(where the variable covers a broad range of scales) 
a very useful procedure is to take a width of the interval $ds$
that is not the same for all $s$,
but that is proportional to the scale,
as $[s,s+ds) =[s_o,b s_o)$, $[b s_o,b^2 s_o)$, $\dots$ $[b^{k} s_o,b^{k+1} s_o)$,
i.e., $ds=(b-1)s$ (with $b>1$).
Given a value of $s$, the corresponding value of $k$
that associates $s$ with its bin is given by
$k=\lfloor \log_b(s/s_o)\rfloor$.
Correspondingly, the optimum choice to assign a point
to the interval $[s,s+ds)$ is given by the value $\sqrt{b} s$.  
This procedure is referred to as logarithmic binning, because the intervals
appear with fixed width in logarithmic scale
\citep{Hergarten_book}.
In this paper we have generally taken $b\simeq 1.58$,
in such a way that $b^5=10$, providing 5 bins per order of magnitude.

As the distributions are estimated from a finite number of data, they display
statistical fluctuations. The uncertainty characterizing these fluctuations
is simply related to the density by 
$$
%\sigma_D(s)/P(s) \simeq 1/\sqrt{n(s)},
\frac{\sigma_P(s)}{P(s)} \simeq \frac 1 {\sqrt{n(s)}},
$$
where $\sigma_P(s)$ is the standard deviation of $P(s)$
(do not confound with the standard deviation of $s$).
This is so because $n(s)$ can be considered a binomial variable \citep{von_Mises}, and then, 
the ratio between its standard deviation and mean fulfills
$\sigma_n(s)/\langle n(s) \rangle 
%%\sout{\sqrt{N_s p (1-p)}/(N_s p) \simeq \sqrt{(1-p)/n(s)}} 
\simeq  1/\sqrt{n(s)}$,
with 
%\sout{$n(s)\simeq p N_s$ and $p \ll 1$}
$n(s)/ N_s \ll 1$.
As $P(s)$ is proportional to $n(s)$, the same relation holds
for its relative uncertainty.

\begin{acknowledgements}
 %\hilightgreen{12: GENERAL REVISION NEEDED. Por mi esta bien.}
This work would not have been possible without the
collaboration of the Ag\`encia Catalana de l'Aigua
(ACA), which generously shared its rain data with us.
We have benefited a lot from a long-term collaboration 
with Ole Peters, 
and are grateful also to R. Romero and A. Rosso, 
for addressing us towards the ACA data and
an important reference \citep{Rosso},
and to J. E. Llebot for providing support and 
encouragement.
We also thank Joaquim Farguell
from ACA.
A. Mugnai and other people of the Plinius conferences
made us realize of the importance of 
our results for the non-linear geophysics community.
A.D. enjoys a Ph.D. grant of the Centre de Recerca Matem\`atica.
%Her initial research was supported from the 
%Explora-Ingenio 2010 project FIS2007-29088-E.
Grants related to this work are FIS2009-09508
and 2009SGR-164.
%%A.C. is a participant of the CONSOLIDER i-MATH project.

\end{acknowledgements}

\bibliographystyle{copernicus}
\bibliography{biblio}

\end{document}